\documentclass[aps,prd,reprint,superscriptaddress,showpacs,nofootinbib]{revtex4-1}

\usepackage{hyperref}
\usepackage{comment}
\usepackage{amsmath}
\usepackage{multirow}
\usepackage{amssymb}
\usepackage{graphicx}
\usepackage{color}
\usepackage{subfig}
\usepackage{bm}
\usepackage{gensymb}
\usepackage{bigints}
\usepackage{placeins}
\usepackage[usenames,dvipsnames,svgnames,table]{xcolor}
\usepackage{pifont}
\usepackage{aas_macros}
\usepackage{orcidlink}
\usepackage{physics}

\newcommand{\Lagr}{\mathcal{L}}

\usepackage[labelfont=bf]{caption}

\hypersetup{
    colorlinks=true,
    linkcolor=Blue,
    filecolor=Blue,
    urlcolor=MidnightBlue,
    citecolor=Blue
}

\numberwithin{equation}{section}

\begin{document}


\title{Massive neutron stars as mass gap candidates: Exploring equation of state and magnetic field}

\author{Zenia Zuraiq\orcidlink{0009-0000-6980-6334}}
\email[E-mail: ]{zeniazuraiq@iisc.ac.in}
\affiliation{Department of Physics, Indian Institute of Science, Bengaluru 560012, India}
\author{Banibrata Mukhopadhyay\orcidlink{0000-0002-3020-9513}}
\email[E-mail: ]{bm@iisc.ac.in}
\affiliation{Department of Physics, Indian Institute of Science, Bengaluru 560012, India}
\author{Fridolin Weber\orcidlink{0000-0002-5020-1906}}
\email[E-mail: ]{fweber@sdsu.edu}
\affiliation{Department of Physics, San Diego State University, 5500 Campanile Drive
San Diego, California 92182, USA
}
\affiliation{Center for Astrophysics and Space Sciences, University of California at San Diego, La Jolla, California 92093, USA}

\begin{abstract}
The densities in the cores of the neutron stars (NSs) can reach several times that of the nuclear saturation density. The exact nature of matter at these densities is still virtually unknown. We consider a number of proposed, phenomenological, relativistic mean field equations of state to construct theoretical models of NSs. We find that, based on our selected set of models, the emergence of exotic matter at these high densities restricts the mass of NSs to $\simeq 2.2 M_\odot$. However, the presence of magnetic fields and a model anisotropy significantly increases the star's mass, placing it within the observational mass gap that separates the heaviest NSs from the lightest black holes. Therefore, we propose that gravitational wave observations, like GW190814, and other potential candidates within this mass gap, may actually represent massive, magnetized NSs.
\end{abstract}

\maketitle

\section{Introduction}
Neutron stars (NSs) are the end products of the stellar evolution of main-sequence stars with masses between $10 \text{ and } 25 M_\odot$. NSs are some of the most extreme objects in the Universe. A typical NS contains a mass on the order of the Sun squeezed into a radius of $\sim 10$ km. At their cores, NSs can have densities several times that of the nuclear saturation density \cite{shapiro, camezind}.

However, there is still much that is unknown about NSs - chief among them being the equation(s) of state (EOS). The strong interactions occurring within the cores of NSs are poorly constrained and, in many cases, virtually unknown. This is further complicated by the fact that, at the exceptionally high densities within NS cores, ``exotic" particles like hyperons and deltas become energetically favorable \cite{Sedrakian:2023PPNP}. Notably, these exotic particles are even less constrained than nucleons themselves and lead to their own theoretical challenges (for a summary of the ``hyperon puzzle" and possible solutions, see \cite{hyppuzzle}). As a result, unlike the celebrated Chandrasekhar limit of white dwarfs, NSs do not have a fixed maximum mass limit (although there have been some EOS-independent estimates - see \cite{tov, ruffini, ns_mass1} for instance). NSs are the only laboratory we know of at present where matter exists in such extreme conditions. As a result, the study of NSs becomes a direct probe of the rich theoretical physics of nuclear matter at high densities.

Theoretically, high-density nuclear matter EOS fall into two main categories - microscopic and phenomenological. The microscopic EOS start from bare two- and three-nucleon interactions to reproduce the nucleon scattering data and properties of bound systems with few nucleons. We will not be focusing on these in the current work. Instead, we explore the phenomenological EOS, which are effective interaction models. A class of phenomenological EOS that has been explored extensively in recent years is the relativistic mean field (RMF) models (for a recent review on NS EOS and associated physics, see \cite{burgio}). These are constructed under the quantum hadrodynamics model, where the interactions in nuclear matter are modeled at the hadronic level, with baryon-baryon interactions taking place through the exchange of mesons. These EOS are constrained in two ways - using the properties of nuclear matter at saturation density, and the observations of NSs themselves.

On obtaining an appropriate EOS, one can construct a family of NSs from it, parametrized by the central density by solving the general relativistic hydrostatic equilibrium equations first proposed by Tolman, Oppenheimer, and Volkoff (TOV) \cite{tov}. Thus, every EOS maps onto a mass-radius (M-R) curve, which is then used to calculate the theoretical mass limit of NSs. Thus, a range of nuclear matter EOS directly results in a range of mass limits.

From an observation perspective, NSs are challenging objects. Although the first radio pulsar observation was made in 1968 \cite{bell}, NS measurements still harbor a lot of uncertainties. Particularly, the radius of a NS is very difficult to measure and is poorly constrained by the available data. Indeed, it is the
measurements of masses, particularly those of heavy NSs, that serve as an important discriminator among the various EOS models.

Observations of ``massive" NSs are, therefore, crucial to constrain the properties of high-density nuclear matter. The heaviest observed NS (as of the time of writing) is the ``black widow" pulsar PSR J0952-0607 \cite{psr4_1, psr4_2}, which has a mass $\rm{M} = 2.35 \pm 0.17$$M_\odot$. On the other hand, the lowest observed black hole mass appears to be around $\rm{M} = 3.3$$M_\odot$ \cite{bh} (however, see also \cite{commentbh}). This has led to an apparent ``mass gap" that exists between the heaviest NSs and the lightest black holes.

Recently, gravitational wave (GW) observations have made significant
strides in addressing this gap \cite{ligo1,ligo2}. In particular, the
GW190814 observation \cite{GW190814} revealed the merger of two
compact objects, one of which was inferred to have a mass within the
range of $2.50 - 2.67 M_\odot$, squarely placing it within the mass
gap. Arguments proposing that this event involved two
black holes have been presented in \cite{Sedrakian:2020PRD}. However, the true
nature of this object, whether it is a black hole or a NS,
remains uncertain, as no electromagnetic counterpart was identified
for this event. Additional observations, such as GW200210-092254
\cite{GW200210}, have also suggested the possibility of objects
residing within the mass gap.

Apart from the possibility of mass gap detections, GW observations also significantly help with constraining NS radii. A parameter that has gained relevance in recent years is the dimensionless tidal deformability ($\Lambda$). This is related to the deformation of the star under external tidal fields, such as that of a companion object in a binary. This deformation leaves an imprint on the GW signal and hence, binary mergers such as GW170817 have been used to place an observational limit on this parameter. Although the limit varies depending on the priors and models considered, it proves to be a useful discriminant between various EOS candidates as it gives a range of acceptable stiffness for a particular EOS model. Since the tidal deformability is closely linked to the radius of the NS, this ends up indirectly providing an additional observational constraint. Combining GW observations with Neutron Star Interior Composition Explorer (NICER) data gives us a radius constraint of around $11 - 13$ km for a $1.4 M_\odot$ NS \cite{nicer1, nicer2}.

As mentioned previously, the heaviest NS has a mass above $2 M_\odot$. Other pulsar observations such as PSR J1614-2230, $\rm{M} = 1.97 \pm 0.04$$M_\odot$  \cite{psr1_1, psr1_2}; PSR J0348+0432, $\rm{M} = 2.01 \pm 0.04$$M_\odot$  \cite{psr2}; and PSR J0740+6620 $\rm{M} = 2.08 \pm 0.07$$M_\odot$  \cite{psr3_1, psr3_2} support the idea that the NS's ``Chandrasekhar" limit (if it exists) could be well above this value. On the other hand, it has been shown through previous work in our group (starting with \cite{kundu}) that the magnetic fields of compact stars can lead to significant enhancements in their masses, both through classical and quantum effects.

NSs are known to have significant magnetic fields. Typical surface fields fall in the range of $10^{8} - 10^{13} \ {\rm G}$. An important subset of NSs is the strongly magnetized ``magnetars," which can have surface fields up to the order of $10^{15} \ {\rm G}$. Magnetars may account for around $10\%$ of the NS population \cite{magnetars, magns}. Emphasizing once more, the field strengths mentioned here are all for the surface values. The fields at the center of the star can end up being orders of magnitude higher than the ones quoted here. 

Hence, it seems that the magnetic fields of NSs and their associated effects on the M-R relationship are not ignorable. The introduction of magnetic fields can affect the star in two ways. Classically, the magnetic field can lead to a magnetic pressure that contributes to the hydrostatic balance of the star. Further, this magnetic pressure can also lead to an overall pressure anisotropy in the NS. Another important effect the magnetic field can have is on the quantum microstates of the nuclear matter itself, i.e., modification of the EOS through Landau quantization. However, as shown previously \cite{sinha}, this effect is only significant for fields above $3 \times 10^{18}\ {\rm G}$. In the present work, we restrict ourselves to fields well below this value and consider only the classical effects of the field.

Along with the anisotropy arising from the magnetic pressure, NSs can have other sources of being anisotropic stars. In a very basic sense, the matter distribution within the star itself can lead to an inherent anisotropy. This can arise due to a variety of physical effects. For instance, convective and/or turbulent mixing in the context of two fluid models can lead to anisotropy \cite{herrera}. At the high densities inside NSs, other effects can contribute to anisotropy. For instance, in the superdense NS cores, the formation of pion/kaon condensates (\cite{pion,kaon}) is favored, leading to pressure reduction along the radial direction, but not overall. The presence of possible superfluidity (\cite{superfluid1, superfluid2}) in NS cores can be an additional source of anisotropy.

On introducing a magnetic field to the star, an additional source of uncertainty arises, which is the magnetic profile within the star. The profile of the magnetic field within the compact star is unknown, which means that one can explore a number of profiles, making sure they are consistent with the Einstein-Maxwell equations. One of the most widely used profiles is a density-dependent exponential profile for the magnetic field magnitude, first proposed by Bandyopadhyay et al. \cite{bandopadhyay}. Although this profile can be shown to be consistent with the Einstein-Maxwell equations (see the discussion in Sec 2.4 of \cite{deb}, for example), some critics have argued that simulations from LORENE indicate the field profile should be more of a polynomial fit \cite{mag_crit,dex1,dex2}. However, one should note that LORENE and hence the resulting polynomial profile are only calculated and inferred in the context of purely poloidal configurations within the star. It has been well known that purely poloidal (or purely toroidal) field profiles in stars lead to instabilities \cite{tayler1,tayler2}, and rather mixed field configurations ensure the most stability, \textbf{e.g. \cite{braithwaite,ciolfi,pili}}. Nevertheless, in this work, as a first approximation, we consider two main orientations for the magnetic field introduced - radially oriented (RO) fields are fields directed toward the radial direction, while transversely oriented (TO) fields are fields directed randomly in a direction perpendicular to the radial direction. The orientation of the magnetic field plays a crucial role. As described in this paper, our approximate choice of model profiles helps to explore a series of realistic EOS in contrast to the other simulations of our group where the field profile is more accurate, but the EOS is approximate \cite{surajit,mayu}. Further, there is no consensus on whether a magnetic field always leads to an increase in the mass of a compact star \cite{cardall, broderick, kayanikhoo}. Thus, the study of magnetized, anisotropic NSs can help resolve multiple open issues.

However, the TOV equations hold for isotropic, spherical stars and are modified in the presence of anisotropy, such as that introduced by magnetic fields. We follow the approach laid out by a previous paper in our group in the same line \cite{deb}, and introduce models for the magnetic field and anisotropy constructed under the assumption of spherical symmetry. This assumption of approximate spherical symmetry is valid in a number of field configurations and magnitudes, as shown previously in our group through two-dimensional simulations \cite{surajit}.

In the present work, we look at the theoretical possibility of massive, magnetized NSs as possible mass gap candidates.  The paper is structured as follows. In Sec. \ref{sec_formalism}, we set up the basic formalism and equations solved to construct our NS models. We describe the EOS used, as well as the models introduced to describe the magnetic field and anisotropy present in the star. In Sec. \ref{sec_res}, we describe our obtained results and discuss possible implications of the same. We explore a different magnetic field profile with its caveats in Sec. \ref{sec_dex}. We end by summarizing our results and with our conclusions in Sec. \ref{sec_con}.


\section{FORMALISM AND SET OF EQUATIONS}
\label{sec_formalism}

To describe NSs, we solve the general relativistic hydrostatic balance equations, i.e., the TOV equations. Because of the introduction of anisotropic effects (both from the magnetic field and general matter effects), the TOV equations are modified. We follow the same modification of the TOV equations outlined in previous work by our group \cite{deb}, given by

\begin{equation} \label{tov1}
    \frac{dm}{dr} = 4 \pi r^2 \left(\rho + \frac{B^2}{8\pi}\right), \   
\end{equation}
\begin{equation} \label{tov2}
    \frac{dp_r}{dr} = 
    \begin{cases}
    \frac{- (\rho + p_r)\frac{\left(4\pi r^3\left(p_r - \frac{B^2}{8\pi}\right) + m \right)}{r(r-2m)} + \frac{2}{r}\Delta}{\left(1 - \frac{d}{d\rho}\left(\frac{B^2}{8\pi}\right)\left(\frac{d\rho}{dp_r}\right)\right)} & \text{(for RO)} \\
    \frac{- (\rho + p_r + \frac{B^2}{4\pi})\frac{\left(4\pi r^3\left(p_r + \frac{B^2}{8\pi}\right) + m \right)}{r(r-2m)} + \frac{2}{r}\Delta}{\left(1 + \frac{d}{d\rho}\left(\frac{B^2}{8\pi}\right)\left(\frac{d\rho}{dp_r}\right)\right)} & \text{(for TO)}.
    \end{cases}
\end{equation}

Here, $m$ denotes the mass, $\rho$ the density, and $B$ the magnitude of the magnetic field at a given radius $r$ within the star. Because of the presence of anisotropy, the pressure along the radial direction, $p_r$ is different from the pressure along the transverse direction $p_t$. This is captured in the effective anisotropy factor $\Delta$ defined as 

\begin{equation} \label{delta}
    \Delta = 
    \begin{cases}
     \left(p_t - p_r + B^2/4\pi\right) & \text{(for RO)} \\
    \left(p_t - p_r - B^2/8\pi\right) & \text{(for TO)}.
    \end{cases}
\end{equation}

The magnetic field thus modifies the expressions for the hydrostatic equilibrium and $\Delta$ in different ways based on its orientation (RO or TO).

\subsection{Modified Bowers-Liang model for anisotropy}

To close the system of equations defined above, we need a model functional form for $\Delta$. We 
use the general parametric form first introduced by Bowers and Liang \cite{bl}. As done previously in our group \cite{deb}, we modify the Bowers-Liang form to further include the effects of the magnetic field. $\Delta$ is then given by

\begin{equation} \label{bl}
    \Delta = 
    \begin{cases}
     \kappa r^2\frac{(\rho+p_r)\left(\rho+3p_r - \frac{B^2}{4\pi}\right)}{1 - \frac{2m}{r}} & \text{(for RO)} \\
    \kappa r^2\frac{(\rho+p_r+\frac{B^2}{4\pi})\left(\rho+3p_r + \frac{B^2}{2\pi}\right)}{1 - \frac{2m}{r}} & \text{(for TO)}.
    \end{cases}
\end{equation}

This model is derived keeping in mind the following key assumptions -
\begin{itemize}
    \item The anisotropic force must vanish at the center, leading to the anisotropy term vanishing quadratically at the center.
    \item Anisotropy varies with position inside the star.
    \item $\Delta$ includes the effects due to local fluid anisotropy as well as the anisotropy due to the magnetic field (both its magnitude and orientation).
\end{itemize}

Following previous work \cite{bl, silva}, we restrict $\kappa$ to the range $[-2/3, 2/3]$. This is to ensure the physicality of the solution and ensure that we do not get a positive $dp/dr$.

We further need to supplement the set of equations (\ref{tov1}, \ref{tov2}, \ref{bl}) with the EOS and magnetic field profile to make it completely solvable.

\subsection{Magnetic field profile}
We introduce a density-dependent magnetic field in the star \cite{bandopadhyay}, given by 
\begin{equation}
\label{mag_bando}
    B(\rho) = B_s + B_0\left[1 - \exp\left\{-\eta{\left(\frac{\rho}{\rho_0}\right)^\gamma}\right\}\right].
\end{equation}

This profile gives us the magnitude of the field as a function of the density, and hence, the radius within the star. Here, $B_s$ corresponds to the surface field of the star, $B_0$ and $\rho_0$ control the field at the center, and $\eta$ and $\gamma$ are model parameters that control how the field decays from center to surface. 

Throughout this work, we have chosen $B_s$ to be $10^{15} \ {\rm G}$. However, the results we obtain are found to be largely independent of this parameter, as long as $B_s$ is not comparable to $B_0$.

\subsection{Equations of state}

To describe the matter present within the star, we use a selection of phenomenological EOS, constructed using the RMF approach. Here, the matter is modeled at the hadron level, with the interactions between the baryons modeled using meson fields as mediators. The meson field strengths are then set to their mean values as per the RMF approximation. Three such meson fields are included in this work - the scalar meson $\sigma$, describing attraction between baryons; the vector meson $\omega$, describing repulsion; and the isovector meson $\rho$, explaining isospin asymmetric interactions.

RMF EOS are constrained in two ways - they must be able to reproduce the observed NS properties, and they must reproduce the properties of symmetric nuclear matter (SNM) at saturation density ($n_0$). 

At the high densities present in NS cores, the presence of exotic particles is energetically favorable. Exotic particles are those that do not exist in stable form under terrestrial conditions. In the present work, we have considered two such classes of exotic particles - hyperons (particles with at least one strange baryon in quark content) and $\Delta$ particles (nonstrange baryons of spin $3/2$). Thus, we explore pure nucleonic ($npe\mu$) EOS along with hyperon and $\Delta$ admixed ($npe\mu-Y\Delta$) EOS.

The general Lagrangian of the relativistic model in the mean field approximation is given by
\begin{equation}
    \Lagr_{RMF} = \Lagr_{baryons} + \Lagr_{mesons} + \Lagr_{leptons}.
\end{equation}

The baryon content ($B$) here comprises the nucleons $N  \in  \{n,p\}$, the hyperons  $Y  \in \{ \Lambda, \Sigma^+, \Sigma^0, \Sigma^-, \Xi^0, \Xi^-\}$ and $\Delta \in \{\Delta^{++}, \Delta^+, \Delta^0, \Delta^- \}$. 
The Lagrangian for interacting baryons is given by
\begin{multline}
    \Lagr_{baryons} = \sum\limits_B \bar{\psi_B} [\gamma_\mu(i\partial^\mu-g_{\omega B}(n)\omega^\mu- \frac{1}{2} g_{\rho B}(n)\mathbf{\tau}.\mathbf{\rho}^\mu)\\  - (m_B - g_{\sigma B}(n)\sigma) ] \psi_B,
\end{multline}
where $g_{\sigma B}(n)$, $g_{\omega B}(n)$, and $g_{\rho B}(n)$ are the meson-baryon coupling constants, $m_B$ is the mass of the baryon, $n$ is the baryon number density, $\mathbf{\tau} = (\tau_1, \tau_2, \tau_3)$ is the Pauli isospin matrix, and $\gamma^\mu$ are the Dirac matrices. The coupling constants are set to reproduce the properties of SNM at $n_0$ within their experimental bounds.

The inclusion of hyperonic matter has been done by including meson-hyperon couplings based on the SU(3) ESC08 model \cite{esc08} and the inclusion of $\Delta$ particles done by including a near-universal meson-$\Delta$ coupling : $x_{i\Delta} = g_{i\Delta}/g_{iN} = 1.2$. This value of $x_{i\Delta}$ is in accordance with the consideration of meson-$\Delta$ coupling based on group theory as explored in recent work \cite{lopes_delta}. Here, $g$ represents the coupling constants; the subscript $\Delta$ indicates the $\Delta$ particles, $N$ the nucleons, and $i$ represents the mesons mentioned above.

The Lagrangians for the leptons ($\in \{e^-, \mu^-\}$) and the mesons are given by

\begin{equation}
    \Lagr_{leptons} = \sum\limits_\lambda \bar{\psi_\lambda} [i\gamma_\mu\partial^\mu - m_\lambda]\psi_\lambda \text{ and}
\end{equation}

\begin{multline}
    \Lagr_{mesons} = \frac{1}{2}(\partial_\mu \sigma \partial^\mu \sigma - m_\sigma^2\sigma^2) - \frac{1}{4}(\omega_{\mu\nu}\omega^{\mu\nu}) + \frac{1}{2}m_\omega^2(\omega_{\mu}\omega^{\mu}) \\ + \frac{1}{2} m_{\rho}^2(\rho_{\mu}\rho^{\mu}) -\frac{1}{4}(\rho_{\mu\nu}\rho^{\mu\nu}).
\end{multline}

Additional nonlinear self-interaction terms (dependent on $\sigma^3$ and $\sigma^4$) are often introduced to ensure that the empirical values of properties like nuclear incompressibility ($K_0$) and effective nuclear mass ($m^*/m_N$) are reproduced correctly. 

Apart from introducing nonlinear terms to the Lagrangian, another approach that ensures the proper reproduction of all saturation properties is to make the coupling constants density dependent. The elimination of the need for nonlinear terms leads to density-dependent RMF parametrizations (DDRMF), introduced as
\begin{equation}
    g_{iB}(n) = g_{iB}(n_0)f_i(x),
\end{equation}
where $i \in\{\sigma, \omega, \rho\}$, $x = n/n_0$. $f(x)$ gives the form of the density dependence, typically described using an ansatz of the form
\begin{equation}
    f_i(x) = a_i\frac{1+b_i(x+d_i)^2}{1+c_i(x+d_i)^2},
\end{equation}

for $i \in \{\sigma, \omega\}$, and
\begin{equation}
    f_\rho(x) = \text{exp}[-a_\rho(x-1)].
\end{equation}

Thus, now, the parameters that must fit to the SNM properties at $n_0$ are the nine density-dependent parameters: $a_\sigma, b_\sigma, c_\sigma, d_\sigma, a_\omega, b_\omega, c_\omega, d_\omega, \text{ and } a_\rho$, along with the meson-baryon couplings at saturation density: $g_{\sigma B}(n_0), g_{\omega B}(n_0), \text{ and } g_{\rho B}(n_0)$.

Another density-dependent modification to the standard RMF scheme comes from the recently constrained slope of the symmetry energy at $n_0$ ($L_0$). Here, the isovector meson-baryon coupling constant alone is modified to take a density-dependent form as
\begin{equation}
    g_{\rho B}(n) = g_{\rho B}(n_0)\text{exp}[-a_\rho(x-1)].
\end{equation}

Following previous work \cite{swl}, we refer to this as the RMFL approach. The value of $L_0$ is then fixed based on the coefficient $a_\rho$ along with the standard meson-baryon coupling constants. 

From the Lagrangian constructed using the RMF/DDRMF/RMFL approach, one can then obtain the baryon and meson field equations by applying the Euler-Lagrange method. As we are constructing RMF models, the meson fields are set to their mean values (values in the ground state). 

Along with the meson mean field equations, we impose charge neutrality and baryon number conservation, leading to a system of five coupled non-linear equations to be solved. Additionally, for the RMFL and DDRMF models, baryon chemical equilibrium has to be imposed as the density-dependent coupling leads to a rearrangement energy contribution (for the detailed systems of equations, see \cite{swl}). The solution of this system of equation gives us the meson mean fields, $\bar{\omega}, \bar{\sigma}, \text{ and } \bar{\rho}$, along with the Fermi momenta of the baryons under consideration \cite{swl}.

We choose a few RMF EOS - GM1L \cite{gm1l}, SWL \cite{swl}, DD2 \cite{dd2}, DD-ME1 \cite{ddme1}, DD-ME2 \cite{ddme2}, and DDMEX \cite{ddmex}- that best satisfy both the constraints from SNM properties and the kind of astrophysical observations we seek to explain in this work. The first two EOS are constructed using the RMFL approach whereas the rest are DDRMF EOS. The SNM properties at $n_0$ for this list of EOS are shown in Table \ref{sat_props_EOS}. 

\begin{table}[!htbp]
\begin{tabular}{ccccccc}
\hline
Property        & GM1L   & SWL    & DD2    & DDME1   & DDME2  & DDMEX   \\ \hline
$n_0~({\rm fm}^{-3})$ & 0.153  & 0.150  & 0.149  & 0.152   & 0.152  & 0.152   \\ 
$E/N~({\rm MeV})$     & -16.3  & -16.0  & -16.02 & -16.20  & -16.14 & -16.097 \\ 
$J~({\rm MeV})$       & 32.5   & 31     & 31.67  & 33.1    & 32.3   & 32.269  \\ 
$L_0~({\rm MeV})$      & 55     & 55     & 55.04  & 55.45   & 51.25  & 49.576  \\ 
$J''~({\rm MeV})$      & -124.6 & -106.2 & -93.23 & -101.05 & -87.19 & -71.47  \\ 
$K~({\rm MeV})$       & 240    & 260    & 242.7  & 244.5   & 250.89 & 267.06  \\
$m_*/m$          & 0.70   & 0.70   & 0.5625 & 0.578   & 0.572  & 0.556   \\ \hline
\end{tabular}

\caption{Properties of SNM at $n_0$ for the EOS used in this work. Properties tabulated here are - energy per nucleon ($E/N$), symmetry energy ($J$), slope of the symmetry energy ($L_0$), the curvature of the symmetry energy ($J''$), the incompressibility ($K$) and the effective mass ($m_*/m$).}

\label{sat_props_EOS}
\end{table}

\begin{figure}[!htpb]
	\centering
	\includegraphics[scale=0.3]{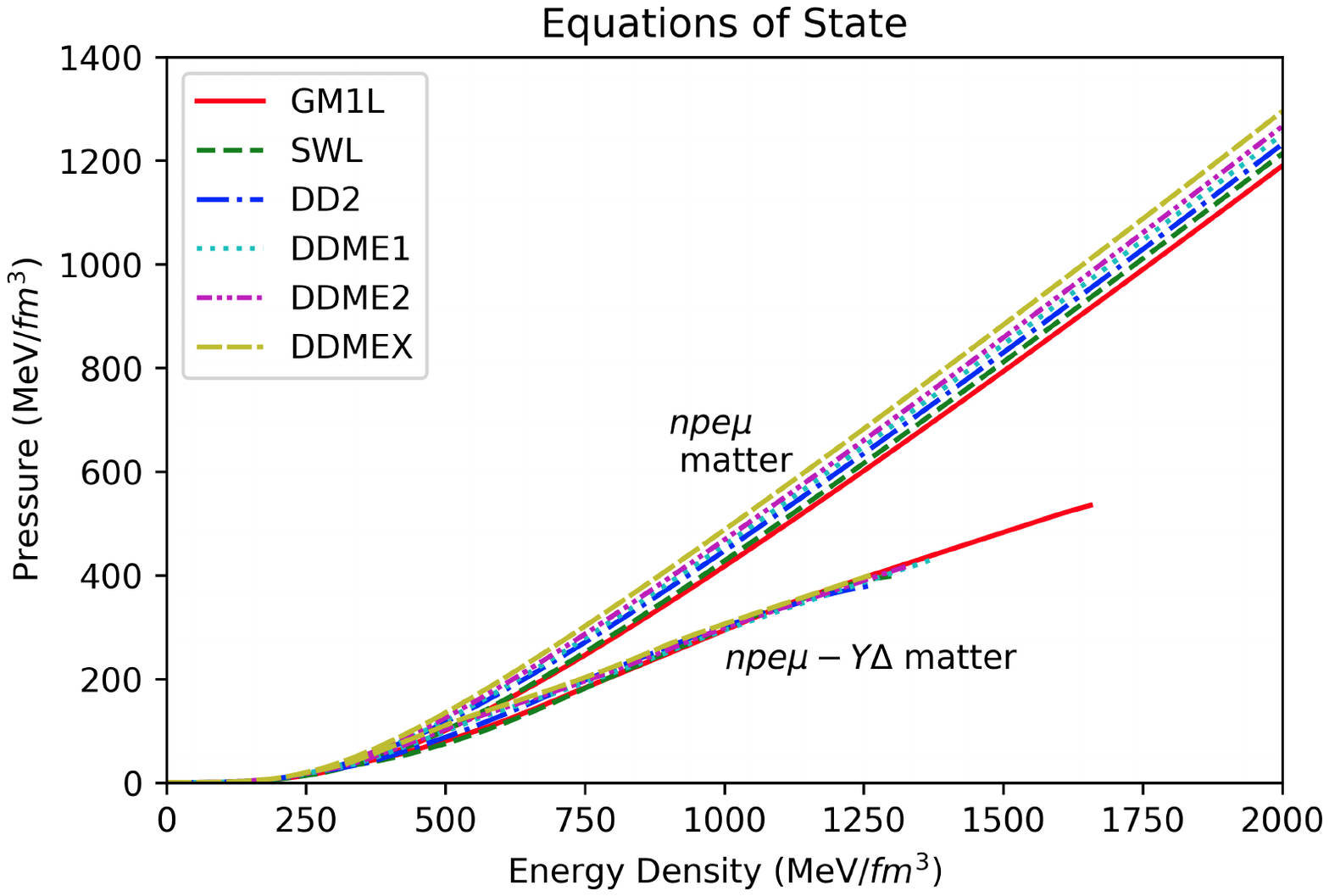}
	\caption{Different RMF EOS: Here, the upper branch denotes the $npe\mu$ EOS, while the lower branch represents the hyperon-$\Delta$ admixed $npe\mu-Y\Delta$ EOS.}
	\label{EOS_plot}
\end{figure}

Figure \ref{EOS_plot} shows that the pure nucleonic EOS are much stiffer (i.e., have a higher pressure at a given energy density) when compared to the cases with exotic particles. This is consistent with past literature. Indeed, the softening of the EOS by hyperonic/exotic matter and the necessity to reconcile with massive NS observations has led to the hyperon puzzle, with different resolutions being proposed, including introducing further repulsive interactions.

\subsection{Tidal deformability}
We can further constrain the NS EOS using tidal deformability limits from GW observations. In the presence of an external gravitational field ($\epsilon_{ij}$), a star develops a quadrupole moment ($Q_{ij}$) such that $Q_{ij} = -\lambda\epsilon_{ij}$, where $\lambda$ is the tidal deformability of the star.

Theoretically, one can link $\lambda$ to the dimensionless second Love number $k_2$, arising from gravitational multipole expansion, as $\lambda = (2/3)k_2\rm{R}^5$ \cite{hinderer}. If $\lambda$ is recast into a dimensionless form, we obtain $\Lambda = \lambda/\rm{M}^5$ $= (2/3)k_2C^{-5}$, where $C$=M/R is the compactness of the star.

We compute the tidal love number $k_2$ by solving the static, linearized perturbation equation arising from the external tidal field. This corresponds to a metric $g_{\alpha\beta} = g_{\alpha\beta}^{(0)} + h_{\alpha\beta}$, where $h_{\alpha\beta}$ is the linearized, perturbation metric. By expanding $h_{\alpha\beta}$ into spherical harmonics, $Y_l^m (\theta, \phi)$, and restricting to the $l = 2$, static, even-parity perturbations in the Regge-Wheeler gauge, we can write \cite{hinderer,biswas}
\begin{multline}
    h_{\alpha\beta} = diag[e^{-\nu(r)}H_o(r), e^{\lambda(r)}H_2(r), \\
     r^2K(r),r^2\text{sin}^2\theta K(r)]Y_{2m}(\theta,\phi),
\end{multline}

\noindent
where $H_0, H_2, \ \text{and } K$ are all radial functions determined by the perturbed Einstein equations. Expanding the perturbed stress-energy tensor and subsequently inserting the fluid and metric perturbations in the linearized Einstein equations, we obtain $H_0 = H_2 \equiv H$ and $K' = H\nu' + H'$. Further subtracting the equation $\delta G_\theta^\theta + \delta G_\phi^\phi = 16\pi\delta p$ from the $tt$ component of the perturbed Einstein equations, we obtain a differential equation for $H$ as
\begin{multline}
    H'' + H'\left[\frac{2}{r} + e^\lambda \left(\frac{2m(r)}{r^2}+4\pi r(p_r-\rho)\right)\right]
    \\ + H\left[4\pi e^\lambda\left(4\rho+8p_r+\frac{\rho+p_r}{Ac_s^2}(1+c_s^2)\right)-\frac{6e^\lambda}{r^2} - \nu^{'2}\right] \\ = 0 \label{love}.
\end{multline} 
Here, $H(r)$ is a radial function arising from the static, linearized perturbations of the Einstein equations. As we are solving for $k_2$, we restrict ourselves to the $l=2$, static, even-parity perturbations of the perturbation metric. The other quantities are $A = dp_t/dp_r$, $c_s^2=dp_r/d\rho$ (the speed of sound squared), $e^\lambda = \left[1-2m/r\right]^{-1}$, and $\nu^{'} = 2e^\lambda(m+4\pi p_r r^3)/r^2$. For isotropic stars, $A = 1$.

On obtaining $H(r)$ by solving Eq. \eqref{love} simultaneously with our system of equations, one can compute the tidal Love number $k_2$ as
\begin{multline}
        k_2 = (8/5)C^5(1-2C^2) \\
        [2-y_R+2C(y_R-1)]\{2C(6-3y_R+3C(5y_R-8)) + 4C^3 \\
        [13-11y_R + C(3y_R-2)+2C^2(1+y_R)]
        +3(1-2C)^2 \\
        [2-y_R+2C(y_R-1)] \text{log}(1-2C)\},^{-1}
\end{multline}\noindent{where $y_R = rH'(\rm{R})/H(\rm{R})$.}

In recent years, due to GW observations, observational constraints have been placed on the dimensionless tidal deformability. From the binary NS merger event GW170817, the dimensionless tidal deformability at $1.4 M_\odot$ has been constrained to be $\Lambda_{1.4} < 800$ \cite{ligo1} and $\Lambda_{1.4} < 580$ \cite{ligo2}. The two different limits are due to different models. We have enforced both in our subsequent analyses.

\section{RESULTS AND DISCUSSION}\label{sec_res}

As mentioned previously, this work investigates the theoretical possibility of NSs with masses high enough to fall in the observational mass gap.

\subsection{Isotropic cases: Pure EOS effect}
\begin{figure}[htpb]
	\centering
	\includegraphics[scale=0.3]{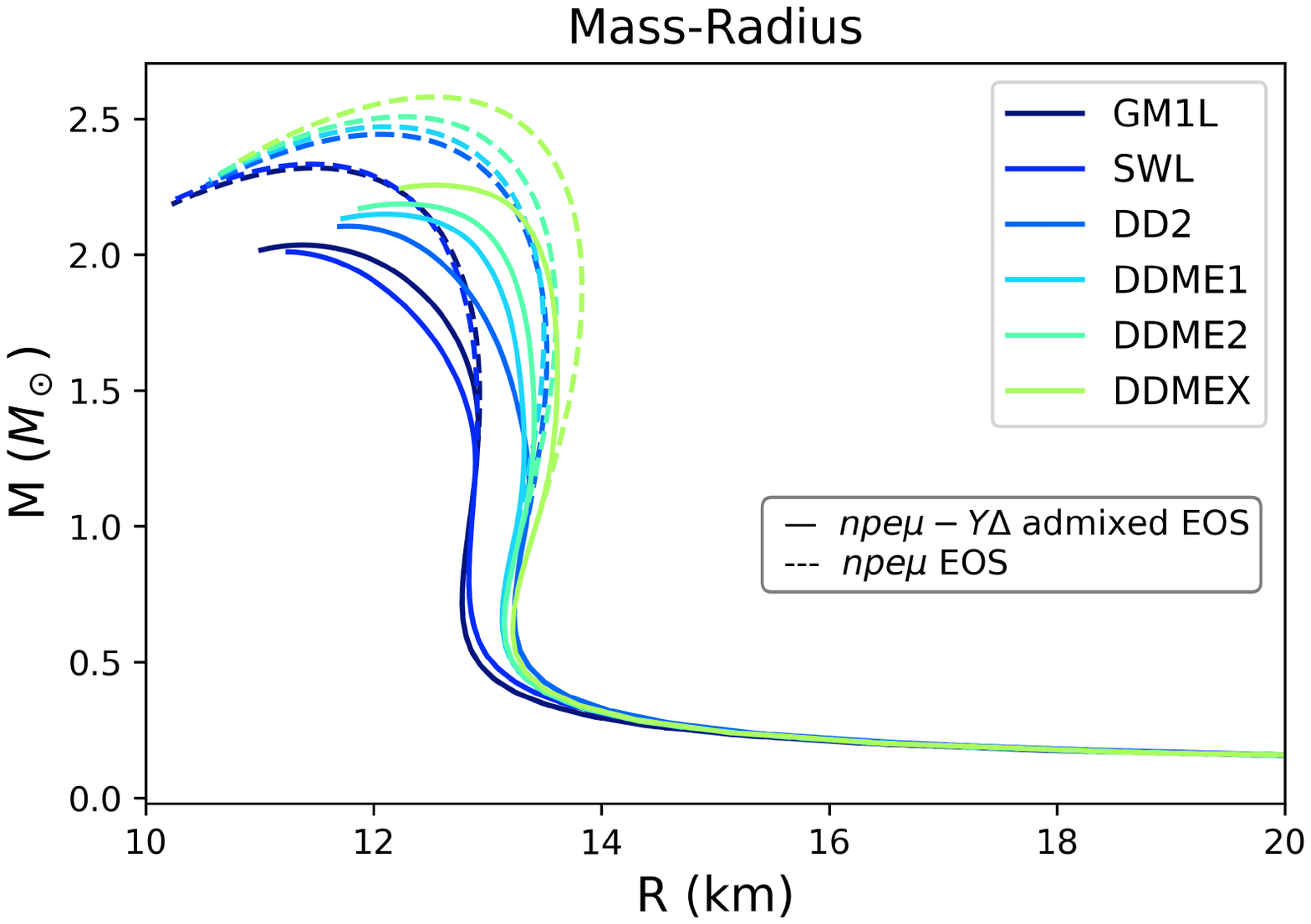}
	\caption{M-R curves for $npe\mu$ EOS (dotted) and $npe\mu-Y\Delta$ EOS (solid). In each set, the top-to-bottom curves are sequentially for DDMEX, DD-ME2, DD-ME1, DD2, SWL, and GM1L.}
	\label{mr_iso_plot}
\end{figure}

To start with, we look at the completely isotropic case - magnetic field and anisotropy parameter $\kappa$ are both set to zero. This gives us an idea of the role that pure EOS effects play in the M-R relationship. M-R curves for the $npe\mu$ and $npe\mu-Y\Delta$ EOS are displayed in Fig. \ref{mr_iso_plot}.

\begin{figure}[htpb]
	\centering
	\includegraphics[scale=0.3]{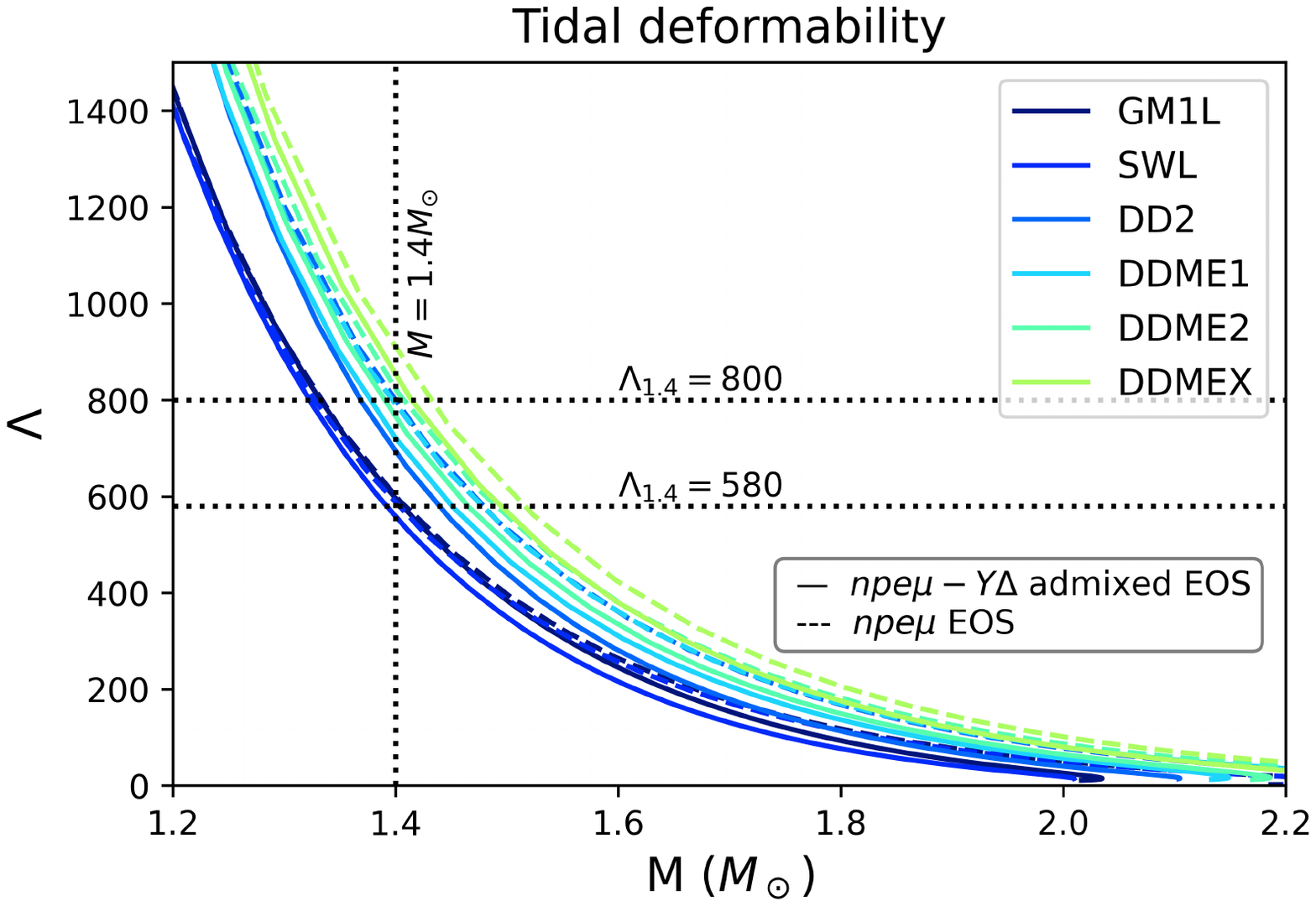}
	\caption{Tidal deformability $\Lambda$ as a function of M for $npe\mu$ EOS (dotted) and $npe\mu-Y\Delta$ EOS (solid). The various curves are the same as Fig. \ref{mr_iso_plot}.}
	\label{tidal_iso_plot}
\end{figure}

As expected, the stiffer $npe\mu$ EOS give higher masses, with the maximum mass limit ($\rm{M}_{max}$) reaching as high as $2.57 M_\odot$ for the DDMEX EOS. The $npe\mu-Y\Delta$ EOS, being softer, give a mass limit $\rm{M}_{max}$ of about $2.25 M_\odot$ (for the same DDMEX EOS). The results for the isotropic cases are shown in Table \ref{iso_table}. However, on computing the tidal deformability (Fig. \ref{tidal_iso_plot}) of these cases, it is clear that the nucleon-only cases all violate at least one of the observational upper bounds, while the hyperon-$\Delta$ admixed cases do not. These results are consistent with NS properties obtained in previous work in this line using the same EOS \cite{swl, eos_verify1, eos_verify2}. Nevertheless, as we will see in further results, the introduction of anisotropy and/or magnetic field helps to change the value of tidal deformability and even bring the nucleon-only cases below observational thresholds. However, based on energy considerations mentioned previously and the tidal deformability constraints, it appears that $npe\mu-Y\Delta$ cases are better favored overall. For the rest of this analysis, we consider only the $npe\mu-Y\Delta$ admixed cases.

\FloatBarrier

\begin{table}[htbp]
\begin{tabular}{ccccc}
\hline
\multirow{2}{*}{EOS} & \multicolumn{2}{c}{$npe\mu$ matter}                                         & \multicolumn{2}{c}{$npe\mu-Y\Delta$ matter}                                    \\ \cline{2-5} 
                     & \multicolumn{1}{c}{$\rm{M}_{max}$ ($M_\odot$)} & \multicolumn{1}{c}{R (km)}  & \multicolumn{1}{c}{$\rm{M}_{max}$ ($M_\odot$)} & \multicolumn{1}{c}{R (km)} \\ \hline
GM1L                 & \multicolumn{1}{c}{2.32}             & \multicolumn{1}{c}{11.45}              & \multicolumn{1}{c}{2.04}             &\multicolumn{1}{c}{11.37}               \\
SWL                  & \multicolumn{1}{c}{2.34}             & \multicolumn{1}{c}{11.43}              & \multicolumn{1}{c}{2.01}             & \multicolumn{1}{c}{11.25}               \\ 
DD2                  & \multicolumn{1}{c}{2.45}             & \multicolumn{1}{c}{11.94}              & \multicolumn{1}{c}{2.11}             & \multicolumn{1}{c}{11.79}                \\ 
DD-ME1                & \multicolumn{1}{c}{2.47}             & \multicolumn{1}{c}{12.12}              & \multicolumn{1}{c}{2.15}             & \multicolumn{1}{c}{12.10}               \\ 
DD-ME2                & \multicolumn{1}{c}{2.51}             & \multicolumn{1}{c}{12.24}               & \multicolumn{1}{c}{2.19}             & \multicolumn{1}{c}{12.25}               \\ 
DDMEX                & \multicolumn{1}{c}{2.58}             & \multicolumn{1}{c}{12.64}               & \multicolumn{1}{c}{2.26}             & \multicolumn{1}{c}{12.53}               \\ \hline
\end{tabular}
\captionsetup{justification=raggedright}
\caption{\label{iso_table}The physical parameters of isotropic NSs, both by inclusion of exotic particles, and for pure nucleonic matter.}
\end{table}
\FloatBarrier

As seen from the isotropic results, pure EOS effects are not sufficient to bring the stable NS to mass gap levels ($> 2.5 M_\odot$). As a result, we turn to the additional physics of magnetic fields and/or anisotropy.

\subsection{Introducing magnetic field with fixed anisotropy parameter, $\kappa$}
\begin{figure}[htpb]
	\centering
	\includegraphics[scale=0.3]{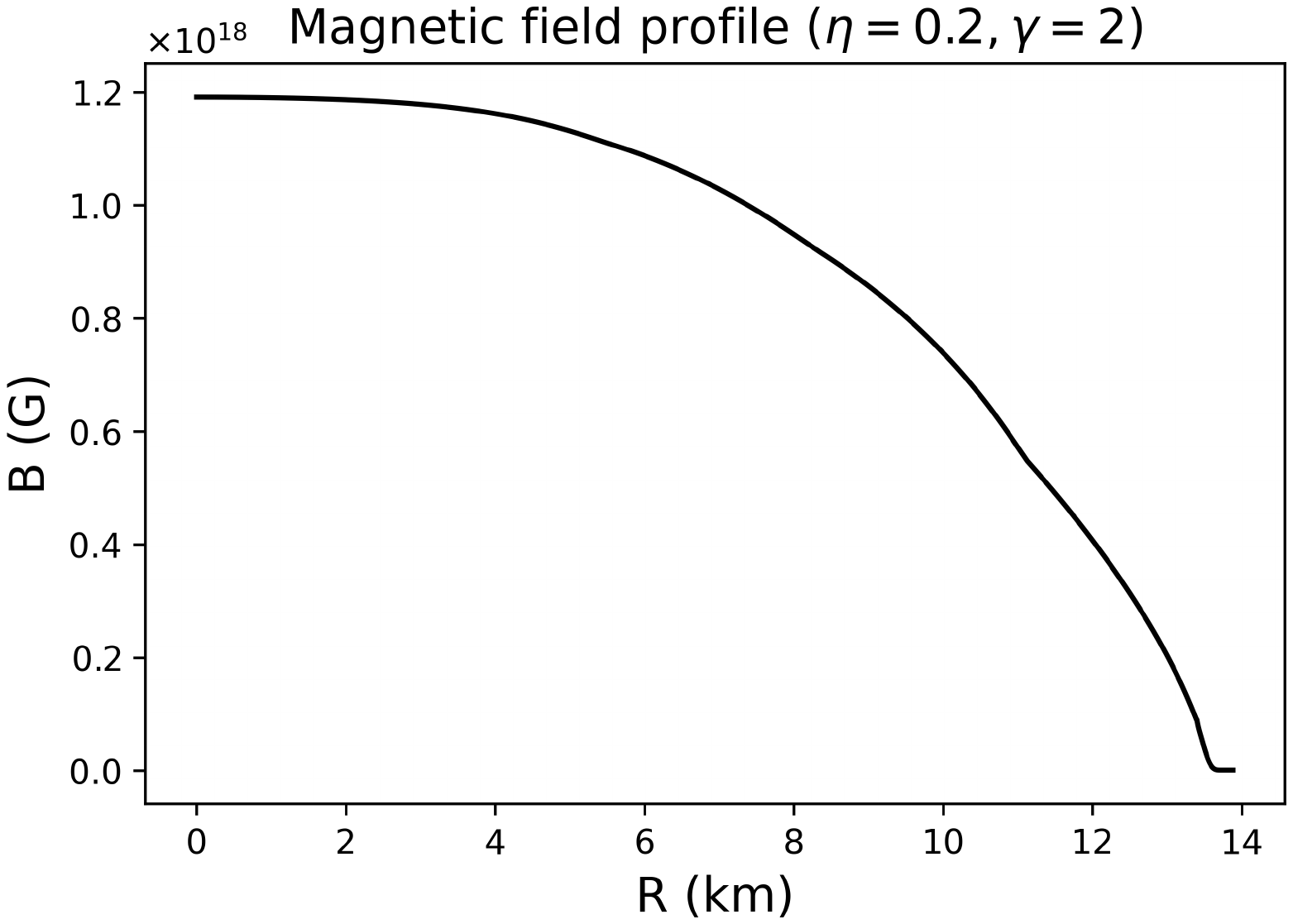}
	\caption{Representative field profile within a star for $\eta = 0.2, \ \gamma = 2$.}
	\label{mag_prof1}
\end{figure}

We first introduce magnetic field following Eq. \eqref{mag_bando} in the presence of a fixed anisotropy parameter, $\kappa = 0.5$. We start with a profile described by $\eta = 0.2, \gamma = 2$. The general shape of this profile within a star is given in Fig. \ref{mag_prof1}. We vary $B_0$ to take values $1.2 \times 10^{18}$ and $0.9 \times 10^{18} \ {\rm G}$ in the transverse direction and $0.6 \times 10^{18}$ and $0.9 \times 10^{18} \ {\rm G}$ in the radial direction. $B_s$ is kept fixed at $10^{15} \ {\rm G}$. As mentioned previously, varying $B_s$ (in a range not comparable to $B_0$) does not have an effect on the results. 

\begin{figure}[!htpb]
	\centering
	\includegraphics[scale=0.3]{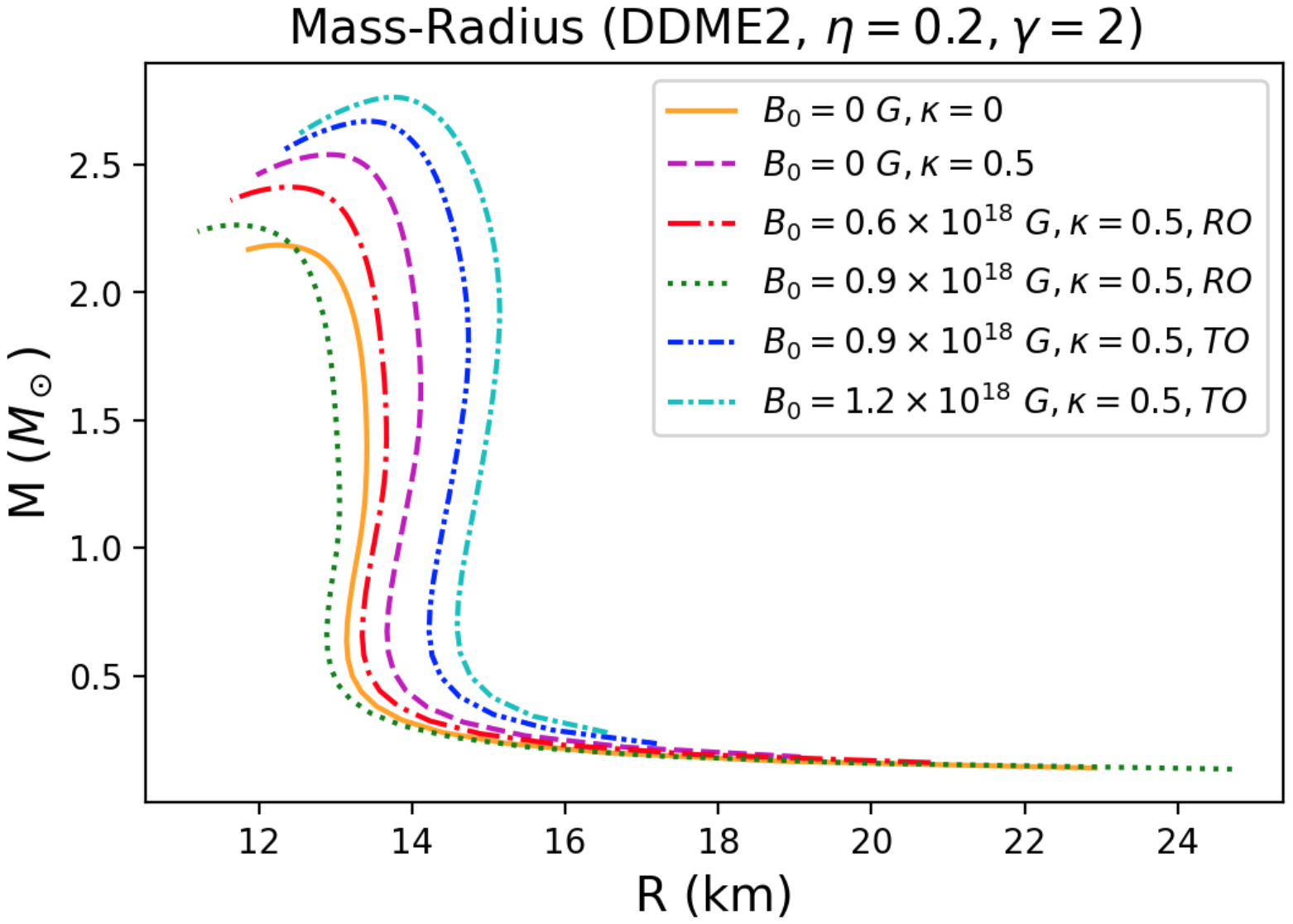}
	\caption{M-R curves for varying $B_0$ and orientation in case of the profile corresponding to $\eta = 0.2, \gamma = 2$ for the DD-ME2 EOS. Anisotropic parameter $\kappa$ is set to $0.5$ throughout.}
	\label{mr_mag1_ddme2}
\end{figure}

\begin{figure}[!htpb]
	\centering
	\includegraphics[scale=0.3]{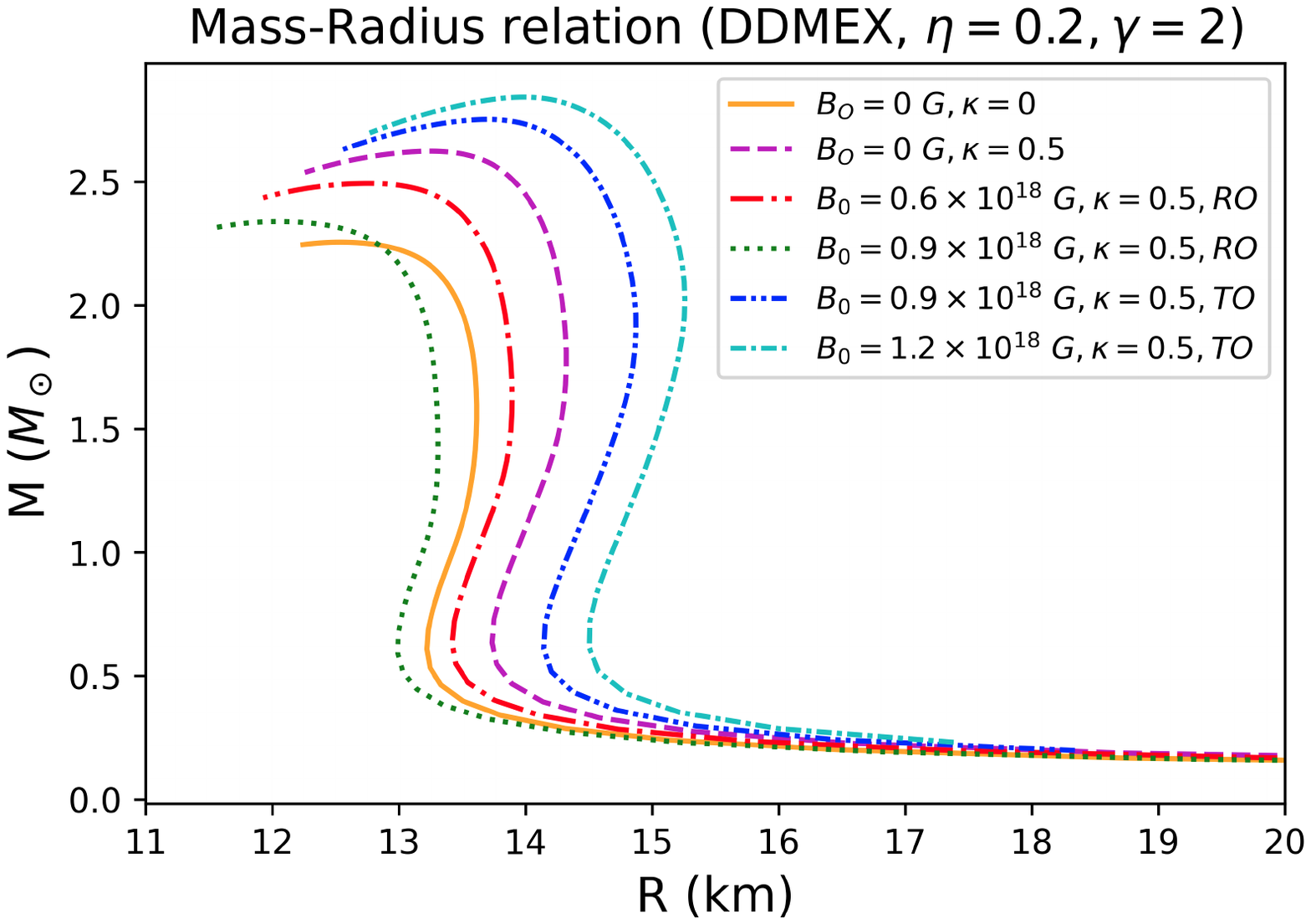}
	\caption{M-R curves for varying $B_0$ and orientation in case of the profile corresponding to $\eta = 0.2, \gamma = 2$ for the DDMEX EOS. Anisotropic parameter $\kappa$ is set to $0.5$ throughout.}
	\label{mr_mag1_ddmex}
\end{figure}

The trend obtained is similar to previous work done in this line \cite{deb}. Introducing anisotropy enhances the mass of the star, even in the absence of magnetic fields. For the latter, it is the matter anisotropy that is ultimately contributing to the enhanced mass of the star. However, it is important to note here that in general anisotropy is not wholly independent of the magnetic field and its effect. On introducing a TO field to an already anisotropic (by matter) star, its mass further increases, whereas a RO field tends to decrease the mass. In these cases, the anisotropy is not wholly a matter effect, but instead has contributions from the magnetic field introduced. This leads to the anisotropy being a consequence and/or an extension of the magnetic effect on the NS's structure and properties. Thus, different M-R curves are obtained on varying the magnetic field. The results for the two stiffest EOS (DD-ME2 and DDMEX) are shown in Figs. \ref{mr_mag1_ddme2} and \ref{mr_mag1_ddmex}. Results for all the EOS are given in Table \ref{mag1_table}. 

\begin{table}[!htbp]
\begin{tabular}{ccccccc}
\hline
\multirow{2}{*}{$B_0~ (10^{18} {\rm G})$} & \multicolumn{6}{c}{$\rm{M}_{max}$ $(M_\odot)$ for different EOS}                                                                                                                      \\ \cline{2-7} 
                                & \multicolumn{1}{c}{GM1L}   & \multicolumn{1}{c}{SWL}    & \multicolumn{1}{c}{DD2}    & \multicolumn{1}{c}{DD-ME1}  & \multicolumn{1}{c}{DD-ME2}  & \multicolumn{1}{c}{DDMEX} \\ \hline
1.2 (TO)                        & \multicolumn{1}{c}{2.57} & \multicolumn{1}{c}{2.52} & \multicolumn{1}{c}{2.64} & \multicolumn{1}{c}{2.72} & \multicolumn{1}{c}{2.76} & 2.84                     \\ 
0.9 (TO)                        & \multicolumn{1}{c}{2.47} & \multicolumn{1}{c}{2.42} & \multicolumn{1}{c}{2.55} & \multicolumn{1}{c}{2.63} & \multicolumn{1}{c}{2.67} & 2.75                     \\ 
0                               & \multicolumn{1}{c}{2.34} & \multicolumn{1}{c}{2.31} & \multicolumn{1}{c}{2.43} & \multicolumn{1}{c}{2.49} & \multicolumn{1}{c}{2.54}  & 2.62                     \\ 
0.6 (RO)                        & \multicolumn{1}{c}{2.24} & \multicolumn{1}{c}{2.21} & \multicolumn{1}{c}{2.32} & \multicolumn{1}{c}{2.37} & \multicolumn{1}{c}{2.41} & 2.49                     \\ 
0.9 (RO)                        & \multicolumn{1}{c}{2.11} & \multicolumn{1}{c}{2.09} & \multicolumn{1}{c}{2.19} & \multicolumn{1}{c}{2.22} & \multicolumn{1}{c}{2.26} & 2.34                     \\ \hline
\end{tabular}
    \caption{$\rm{M}_{max}$ for the different EOS under varying $B_0$ and orientations. Anisotropy parameter $\kappa = 0.5$ throughout. The field profile is for $\eta = 0.2, \gamma = 2$. $\kappa$ by itself leads to higher masses, which are further enhanced (reduced) by TO (RO) fields.}
    \label{mag1_table}
\end{table}

\begin{figure}[!htpb]
	\centering
	\includegraphics[scale=0.5]{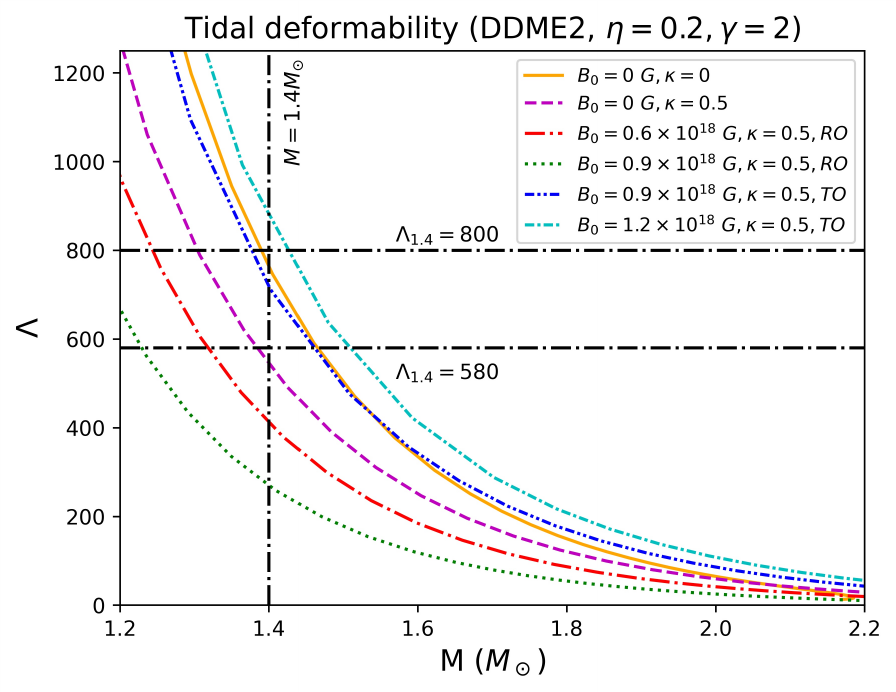}
	\caption{Tidal deformability $\Lambda$ as a function of M for varying $B_0$ in case of the profile corresponding to $\eta = 0.2, \gamma = 2$ for the DD-ME2 EOS. Anisotropic parameter $\kappa$ is set to $0.5$ throughout.}
	\label{tidal_mag1_ddme2}
\end{figure}

\begin{figure}[!htpb]
	\centering
	\includegraphics[scale=0.3]{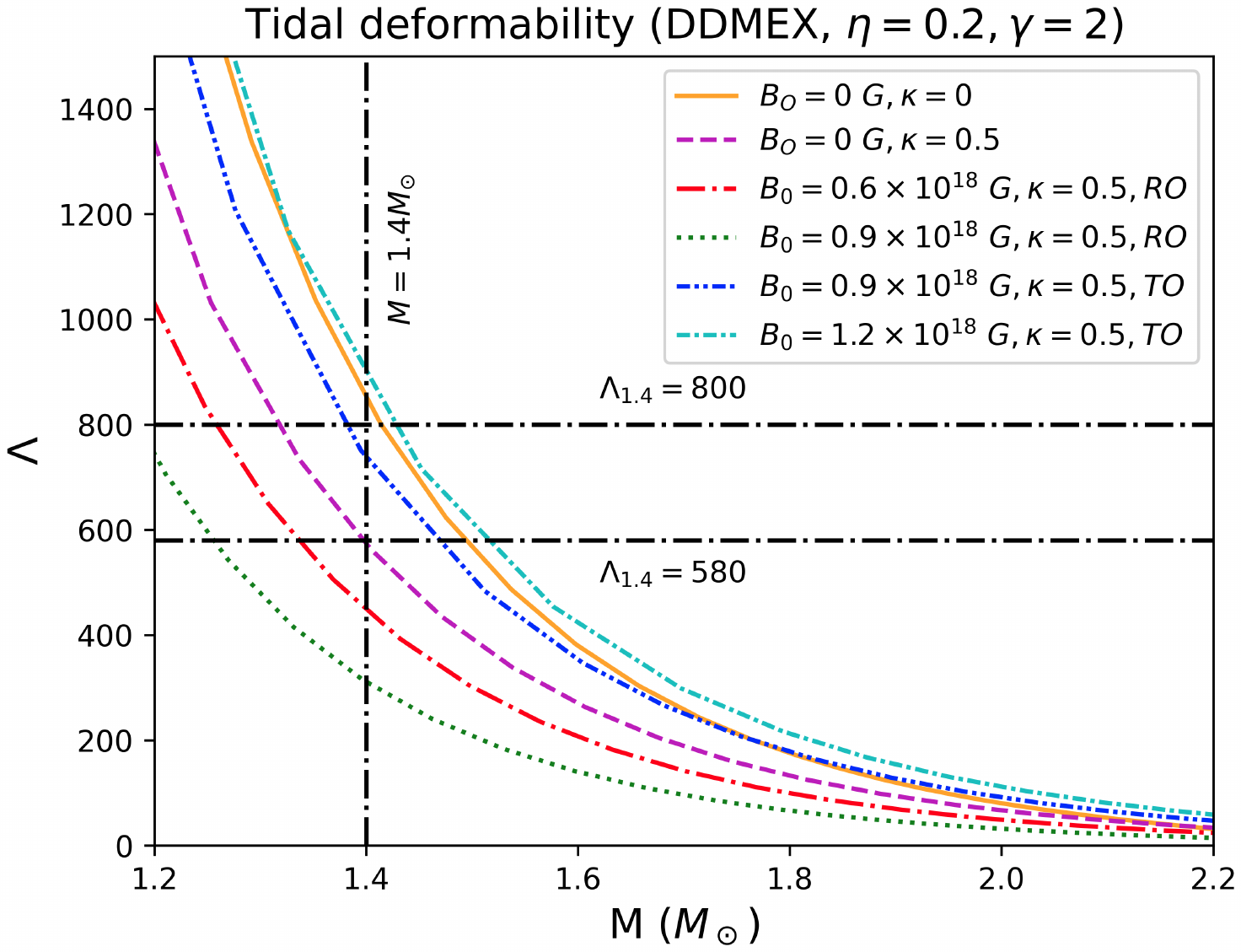}
	\caption{Tidal deformability $\Lambda$ as a function of M for varying $B_0$ in case of the profile corresponding to $\eta = 0.2, \gamma = 2$ for the DDMEX EOS. Anisotropic parameter $\kappa$ is set to $0.5$ throughout.}
	\label{tidal_mag1_ddmex}
\end{figure}

It seems that the highly magnetized TO stars are most promising in terms of mass gap candidates. As seen from Table \ref{mag1_table}, all the EOS give stars of maximum mass $>2.5 M_\odot$ for the TO field with $B_0 = 1.2 \times 10^{18} \ {\rm G}$. The mass of the NS is increased up to even $2.8 M_\odot$ for the DDMEX EOS. However, on computing the tidal deformability for these cases (Figs. \ref{tidal_mag1_ddme2} and \ref{tidal_mag1_ddmex}), we see that these highly magnetized cases fail to meet either of the observational bounds set by GW observations. Similar issues arise when we perform a stability analysis \cite{braithwaite}, as all the $B_0 = 1.2 \times 10^{18}$ G cases give high magnetic energy ($E_{mag}$) to gravitational energy ($E_{grav}$) ratios of order 0.1 (not explicitly shown here).

\begin{figure}[!htpb]
	\centering
	\includegraphics[scale=0.5]{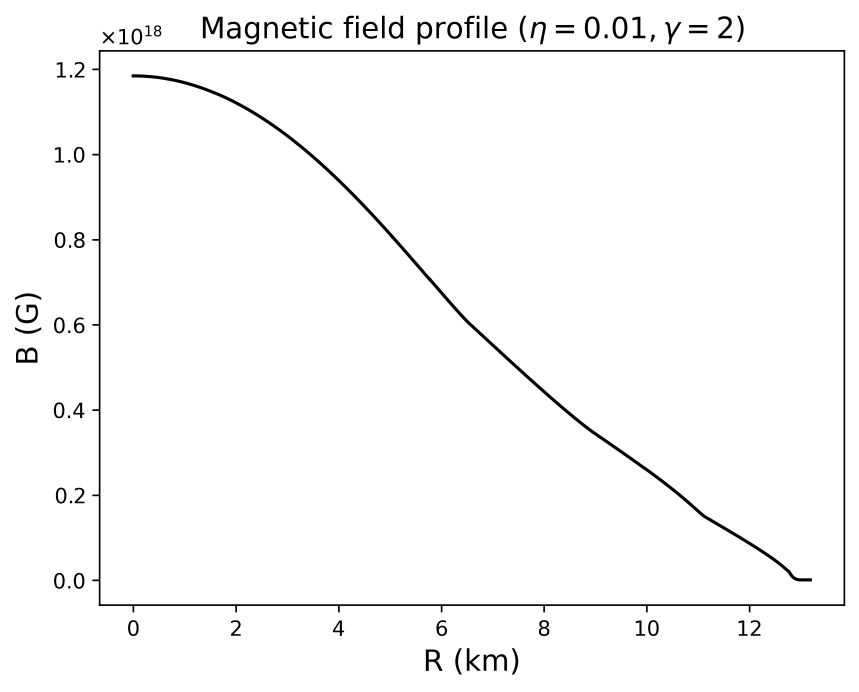}
	\caption{Representative field profile within a star for $\eta = 0.01, \ \gamma = 2$.}
	\label{mag_prof_2}
\end{figure}

Can we then rule out this much high fields at the cores of NSs completely? The answer is \textit{no}. It is important to note that the results discussed so far are profile specific - i.e., specific to a particular $\eta$ and $\gamma$. We next try a second profile: $\eta = 0.01, \gamma = 2$. The general shape of this profile within the star is shown in Fig. \ref{mag_prof_2}. On comparing with the previous profile, we see that the magnetic field here has a much more gradual fall in the star, leading to a lower Lorentz force near the surface of the star. Hence, we expect that this profile will lead to lower masses, however, with better prospects for stability.

\begin{figure}[!htpb]
	\centering
	\includegraphics[scale=0.3]{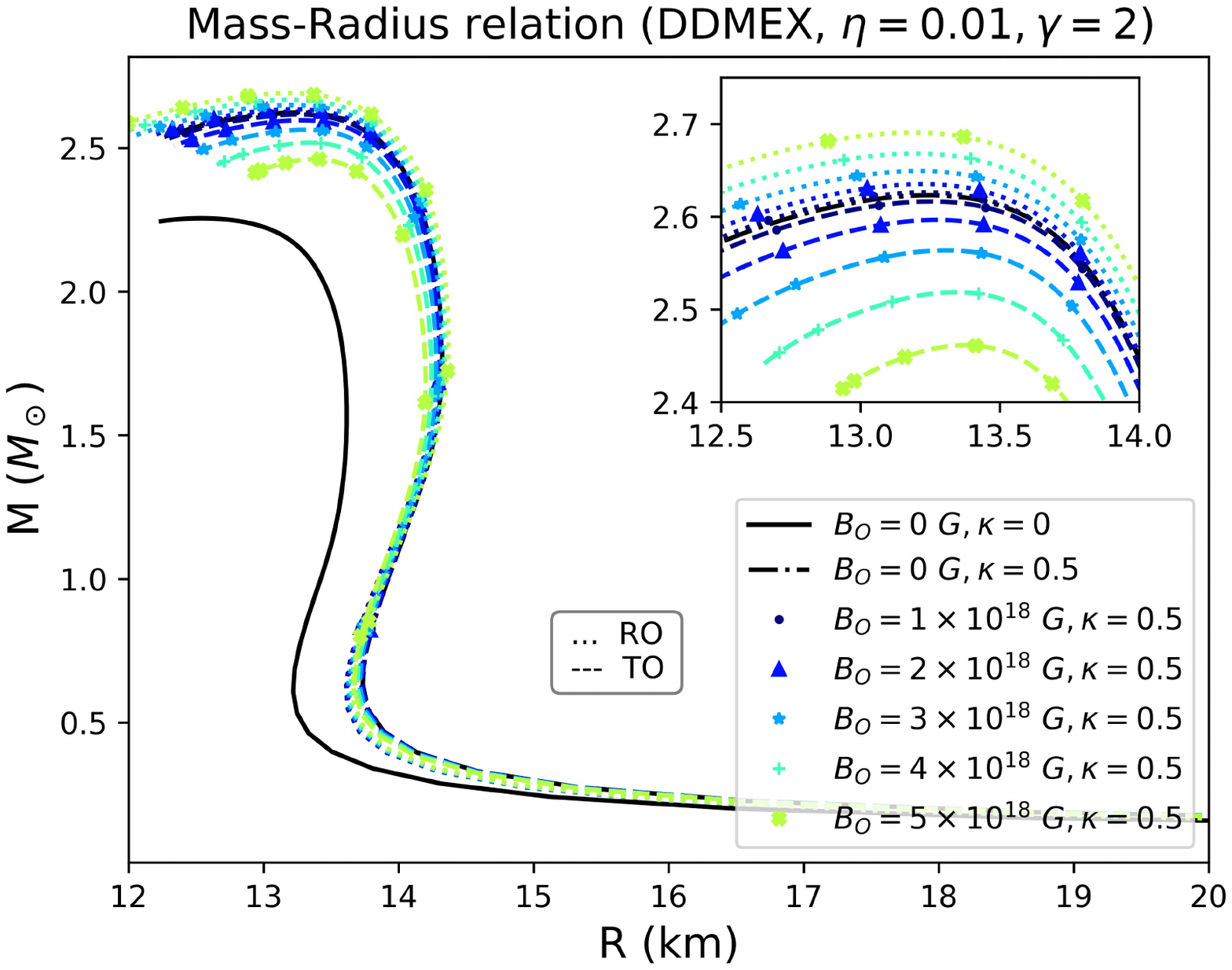}
	\caption{M-R curves for varying $B_0$ in case of the field profile with $\eta = 0.01, \gamma = 2$ for the DDMEX EOS. Anisotropic parameter, $\kappa$, is set to $0.5$ throughout. Dotted lines indicate RO fields, dashed lines indicate TO fields.}
	\label{mr_mag_2}
\end{figure}

We introduce RO and TO fields by varying $B_0$ to take the values - $1 \times 10^{17}$, $5 \times 10^{17}$, $1 \times 10^{18}$, $2 \times 10^{18}$, $3 \times 10^{18}$, $4 \times 10^{18}$, and $5 \times 10^{18}$. The M-R curves obtained in the case of the DDMEX EOS are shown in Fig \ref{mr_mag_2}. The important quantities are also listed in Table \ref{mag2_table}.

\begin{table}[!htbp]
\begin{tabular}{ccccc}
\hline
$B_{0} \ (10^{18}\ {\rm G})$ & $B_{c} \ (10^{17} \ {\rm G})$ & $\rm{M}_{max}$ $(M_\odot)$ & R (km) & $E_{mag}/E_{grav}$ \\ \hline
$5$ (RO)             & 10.71 (RO)           & 2.46              & 13.39     & 0.044            \\ 
$4$ (RO)             & 9.19 (RO)           & 2.52              & 13.33     & 0.031            \\ 
$3$ (RO)             & 7.14 (RO)           & 2.56              & 13.30     & 0.019             \\ 
$2$ (RO)             & 4.87 (RO)           & 2.59              & 13.27     & 0.009            \\ 
$1$ (RO)             & 2.47 (RO)           & 2.61              & 13.25     & 0.002            \\ 
$0.5$ (RO)           &  1.23 (RO)                      & 2.62              & 13.24     &   0.0005                 \\ 
$0.1$ (RO)           &   0.25 (RO)                   & 2.62              & 13.26     &      $3 \times 10^{-5}$              \\ 
$0$                  & $0$                   & 2.62              & 13.26     & -                  \\ 
$0.1$ (TO)           &   0.25 (TO)                    & 2.62              & 13.26     &     $3 \times 10^{-5}$               \\ 
$0.5$ (TO)           &    1.24 (TO)                   & 2.62              & 13.25     & 0.0006                   \\ 
$1$ (TO)             & 2.46 (TO)           & 2.63              & 13.25     & 0.002            \\ 
$2$ (TO)             & 4.92 (TO)           & 2.64              & 13.23     & 0.009            \\ 
$3$ (TO)             & 7.38 (TO)           & 2.65              & 13.21    & 0.02           \\ 
$4$ (TO)             & 9.50 (TO)           & 2.67              & 13.21     & 0.04            \\ 
$5$ (TO)             & 11.89 (TO)           & 2.69             & 13.18     & 0.056            \\ \hline
\end{tabular}
\caption{Results for the DDMEX EOS under varying $B_0$ and orientations. Anisotropy parameter $\kappa = 0.5$ throughout. The field profile is for $\eta = 0.01, \gamma = 2$.}
    \label{mag2_table}
\end{table}

As seen from the results, we require central fields ($B_c$) of around $5 \times 10^{17} \ {\rm G}$ or higher for the magnetic field to significantly affect the mass of the star. For comparable values of $B_c$ in both chosen profiles, we see that the second profile leads to lower masses. For DDMEX, for instance, the second profile gives a maximum mass of $2.69 M_\odot$ as opposed to $2.84 M_\odot$ in the first profile. However, $E_{mag}/E_{grav}$ is now reduced from nearly $0.2$ for the first profile to $0.05$ for the second. Similar trends are seen for the other EOS used. Thus, this second profile gives us much more stable stars overall.

\begin{figure}[!htpb]
	\centering
	\includegraphics[scale=0.6]{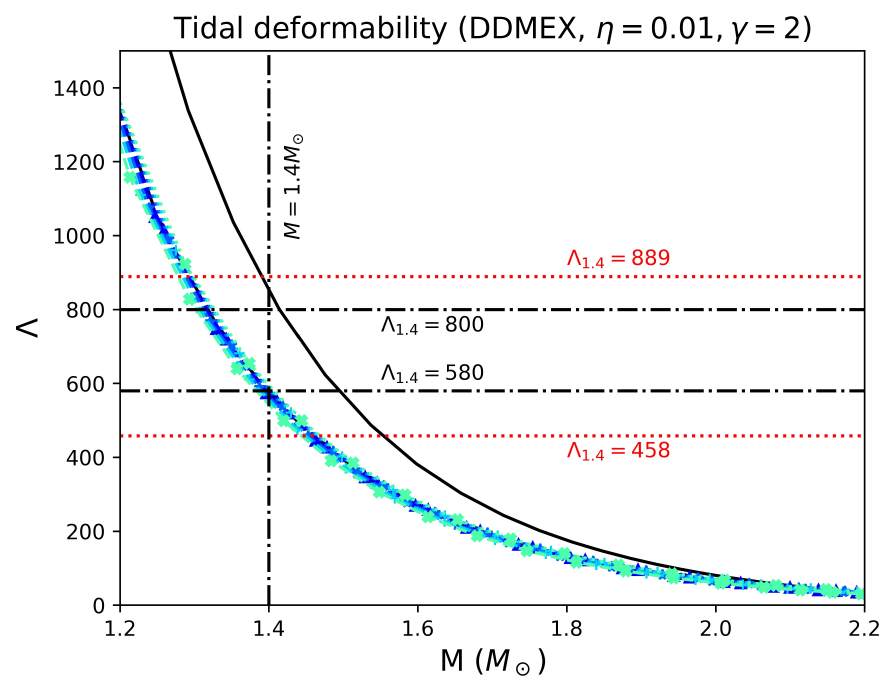}
	\caption{Tidal deformability $\Lambda$ as a function of M for varying $B_0$ in case of profile corresponding to $\eta = 0.01, \gamma = 2$ for the DDMEX EOS. Anisotropic parameter $\kappa$ is set to $0.5$ throughout. The red, dotted lines show the limits on $\Lambda_{1.4}$ from GW190814. The various curves are the same as Fig. \ref{mr_mag_2}.}
	\label{ddmex_tidal_mag2}
\end{figure}

We further look at the tidal deformability computed for the second profile. The results are shown in Fig. \ref{ddmex_tidal_mag2}. We see that the lower $\eta$ leads to much lower $\Lambda_{1.4}$. In fact, most of the values of $\Lambda_{1.4}$ are clustered around $580$. This means that these stars are better candidates to satisfy even the stricter observational limit on $\Lambda_{1.4}$.

Although there was no measurable tidal deformations in the GW190814 signal, there have been tidal deformability constraints from this event, obtained from reexamining the analysis of GW170817 under the assumption of GW190814 having a massive NS. In particular, the spectral EOS distribution from GW170817 was taken and each EOS was reweighted by the probability that its maximum mass falls in the mass of the secondary object detected in the GW190814 \cite{GW190814}. This favors stiffer EOS with the updated tidal deformability constraint being $458 < \Lambda_{1.4} < 889$. These updated limits are additionally shown in Fig. \ref{ddmex_tidal_mag2}. We see that all the cases, including the highly magnetized ones, are consistent with all the tidal deformability constraints laid out so far. 

Exploring just these two profile cases, we thus see a wide variety of behavior possible by experimenting with the model parameters of the magnetic field. Since the true form of the magnetic field within the star is essentially unknown, it is important to experiment with different profiles and check the results. In particular, we cannot rule out the possibility of high fields existing within NSs based on results from one profile alone.

\subsection{Varying the anisotropy parameter, $\kappa$}
\begin{figure}[!htpb]
	\centering
	\includegraphics[scale=0.3]{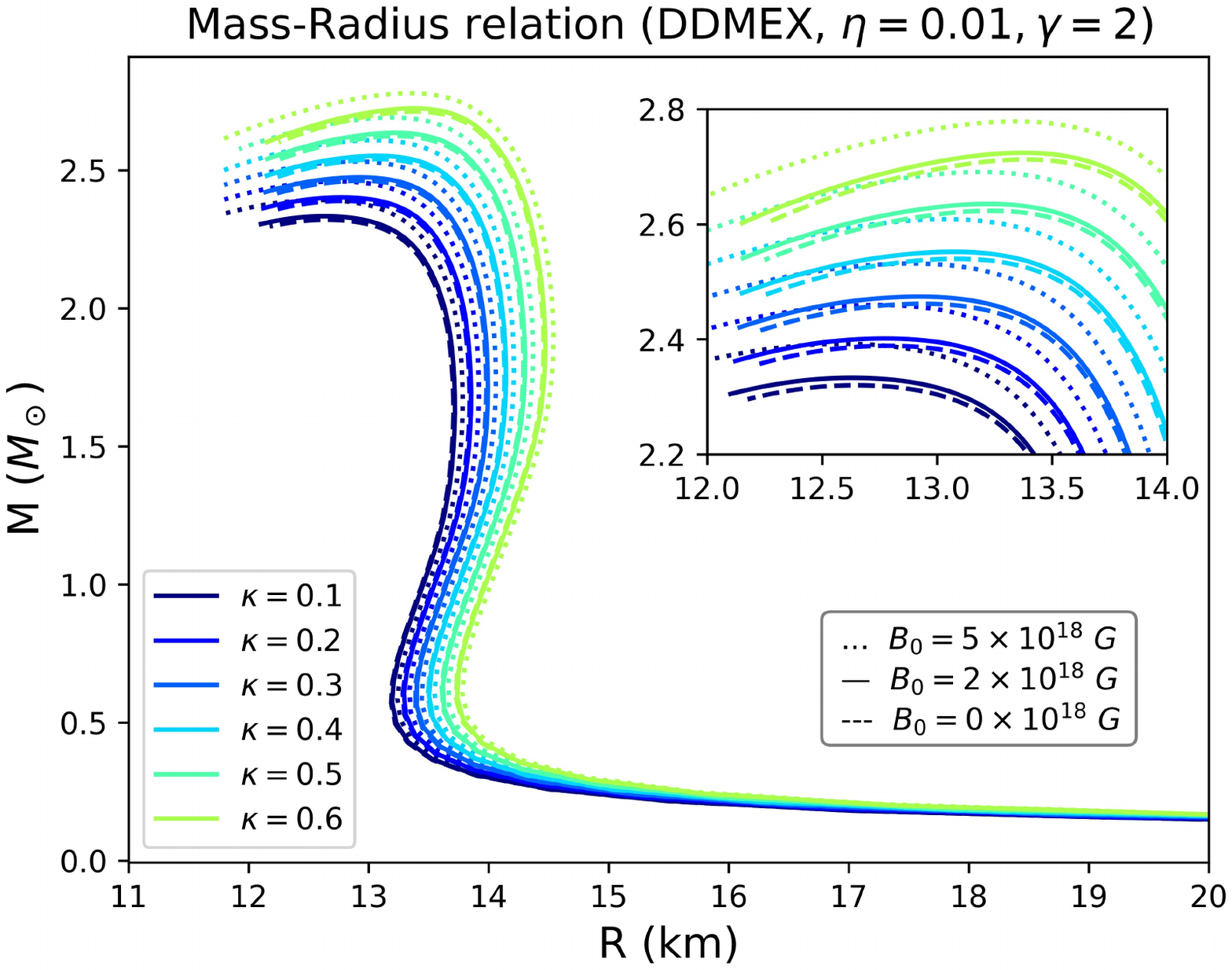}
	\caption{Change in the M-R curve of the DDMEX EOS due to varying anisotropy parameter $\kappa$. Three different $B_0$ have been considered: $0$, $2 \times 10^{18}$, and $5 \times 10^{18} \ {\rm G}$. From top to bottom, each set corresponds to $\kappa = 0.6, \ 0.5, \ 0.4, \ 0.3, \ 0.2$, and $0.1$ sequentially.}
	\label{change_kappa}
\end{figure}

\begin{figure}[!htpb]
	\centering
	\includegraphics[scale=0.3]{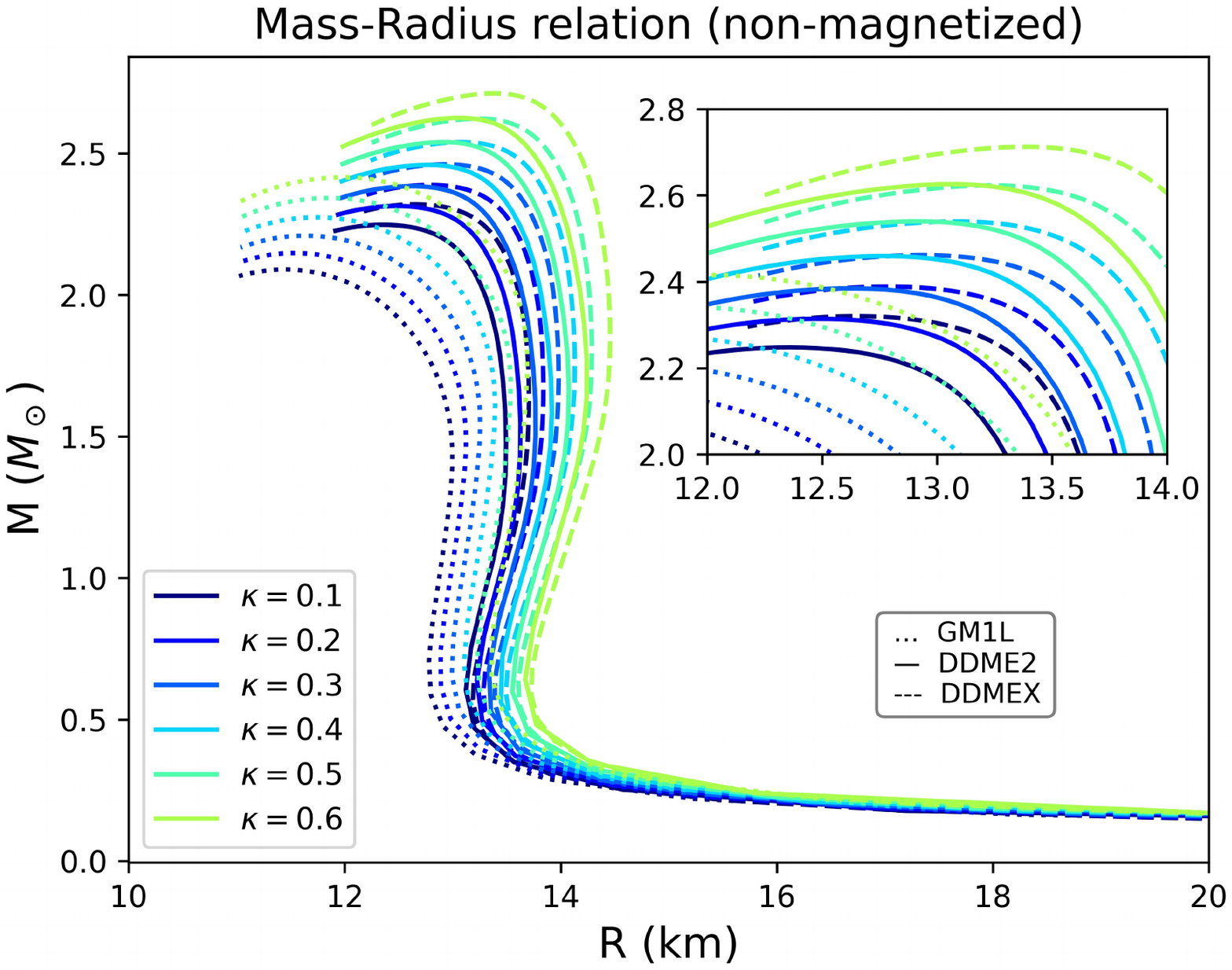}
	\caption{Change in the M-R curve for EOS GM1L, DD-ME2, and DDMEX due to varying anisotropy parameter $\kappa$. $B_0$ has been fixed to $0 \ {\rm G}$ throughout. From top to bottom, the curves for each EOS correspond to $\kappa = 0.6, \ 0.5, \ 0.4, \ 0.3, \ 0.2$, and $0.1$ sequentially.}
	\label{change_kappa_EOS}
\end{figure}

We next look at the effect of varying $\kappa$ in the presence of a fixed magnetic field. We take $\kappa$ as $0.1$, $0.2$, $0.3$, $0.4$, $0.5$, and $0.6$ by fixing $B_0$. In Fig. \ref{change_kappa}, we show the M-R curves for varying $\kappa$ with three fixed $B_0$: $0$, $2 \times 10^{18}$, and $5 \times 10^{18} \ {\rm G}$, for the DDMEX EOS. In Fig. \ref{change_kappa_EOS}, the effect of changing $\kappa$ is shown for three different EOS: GM1L, DD2, DDMEX, at $B_0 = 0 \ {\rm G}$ (i.e., nonmagnetized star).

\begin{figure}[!htpb]
	\centering
	\includegraphics[scale=0.3]{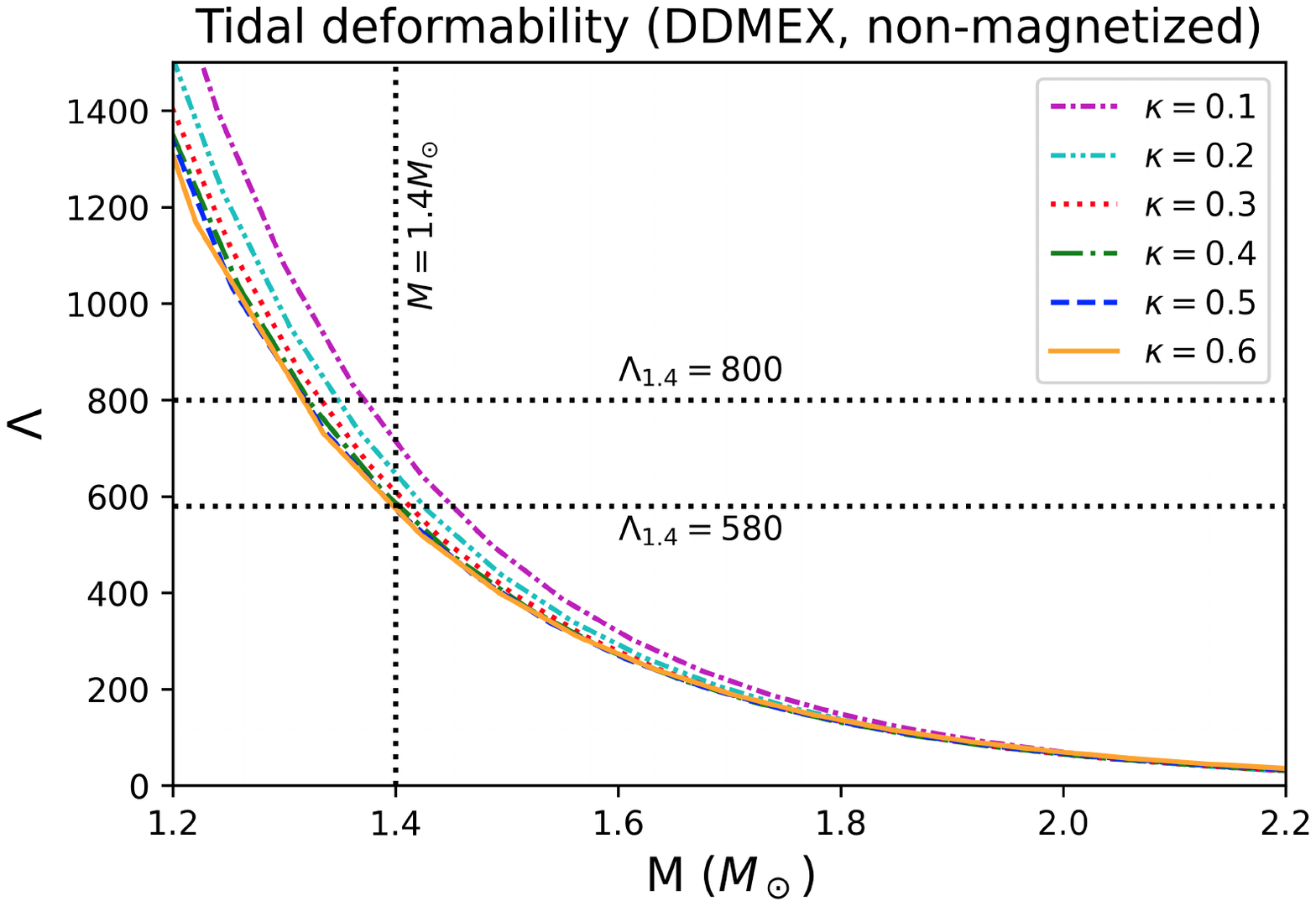}
	\caption{Change in $\Lambda-M$ relation of the DDMEX EOS due to varying anisotropy parameter $\kappa$. $B_0$ has been fixed to $0 \ {\rm G}$ throughout.}
	\label{tidal_kappa}
\end{figure}

We see that higher values of $\kappa$ lead to higher masses being supported by the star, even at zero field. From the results tabulated in Table \ref{vary_kappa_table}, we see that for a given field, a higher $\kappa$ leads to a lower $E_{mag}/E_{grav}$ (and lower $B_c$) along with a higher mass. Higher $\kappa$ also reduces the tidal deformability (Fig. \ref{tidal_kappa}). EOS that violate the observational bounds on $\Lambda$ in the case of isotropic cases can be seen to be consistent with the same bounds when we introduce anisotropy to the star. This dependence of $\Lambda$ on the degree of anisotropy present in the star could be crucial in determining the observational bounds on $\kappa$. Indeed, previous work \cite{biswas} discussed how $\kappa$ could be eventually constrained using GW observations. 

\begin{table}[!htbp]
\begin{tabular}{cccccc}
\hline
$B_0 (G)$                                  & $\kappa$ & $B_c (10^{17} \ G)$ & $\rm{M}_{max}$ $(M_\odot)$ & R (km) & $E_{mag}/E_{grav}$ \\ \hline
\multicolumn{1}{c}{\multirow{6}{*}{$0$}} & 0.1      & -                  & 2.32   & 12.65   & -                  \\ 
\multicolumn{1}{c}{}                     & 0.2      & -                  & 2.39    & 12.79   & -                  \\  
\multicolumn{1}{c}{}                     & 0.3      & -                  & 2.46     & 12.95   & -                  \\  
\multicolumn{1}{c}{}                     & 0.4      & -                  & 2.54    & 13.08   & -                  \\  
\multicolumn{1}{c}{}                     & 0.5      & -                  & 2.62    & 13.25   & -                  \\ 
\multicolumn{1}{c}{}                     & 0.6      & -                  & 2.71    & 13.39   & -                  \\ \hline
\multirow{6}{*}{$2 \times 10^{18}$}        & 0.1      & 6.65               & 2.33   & 12.64   & 0.013             \\ 
                                           & 0.2      & 6.29               & 2.40    & 12.76   & 0.012             \\ 
                                           & 0.3      & 5.76               & 2.47    & 12.93   & 0.011             \\ 
                                           & 0.4      & 5.42               & 2.55    & 13.06   & 0.010             \\ 
                                           & 0.5      & 4.92               & 2.63    & 13.23   & 0.009            \\  
                                           & 0.6      & 4.44               & 2.72    & 13.37   & 0.008            \\ \hline
\multirow{6}{*}{$5 \times 10^{18}$}        & 0.1      & 15.7               & 2.39    & 12.59   & 0.08              \\  
                                           & 0.2      & 14.8               & 2.46    & 12.72   & 0.07              \\ 
                                           & 0.3      & 13.5               & 2.53   & 12.90   & 0.064             \\ 
                                           & 0.4      & 12.6               & 2.61    & 13.04  & 0.06             \\ 
                                           & 0.5      & 11.9               & 2.69    & 13.18   & 0.056            \\  
                                           & 0.6      & 11.8               & 2.78    & 13.33   & 0.052             \\ \hline
\end{tabular}

\caption{Results for the DDMEX EOS under varying $\kappa$ for three $B_0$: $0$, $2 \times 10^{18}$, and $5 \times 10^{18} \ {\rm G}$. The profile corresponds to $\eta = 0.01, \gamma = 2$.}
    \label{vary_kappa_table}
\end{table}

However, as of now, there are not any observational bounds on $\kappa$. Nevertheless, the presence of anisotropy within a NS is well motivated, both by the possible presence of a magnetic field and by high-density effects, such as superfluidity. Previous studies also explored the possibility of the mass gap object in GW190814 being an anisotropic NS \cite{GW190814_aniso}.


\subsection{Universal relations}
As mentioned throughout this paper, the primary challenge arising in the theoretical study of NSs is the unknown EOS. However, unlike the M-R curves explored so far, there are ``universal relations" between NS parameters that are independent/insensitive to the underlying EOS. Such relations are found to exist between the moment of inertia ($I$) of the star, stellar compactness ($C$), and the tidal deformability (Love). This leads to the $I$-Love-$C$ relations. Replacing $C$ with the quadrupole moment $Q$ leads to the other famous set of universal relations, the $I$-Love-$Q$ relations. These universal relations exhibit immediate applicability to various realms of physics. They help to break the degeneracy that exists between the quadrupole moment and spin in GW signals. They can also be a source of EOS-independent tests of the fundamental physics of general relativity \cite{ilovec, iloveq}.

These universal relations were first introduced/studied in the context of isotropic NSs constructed under the slow rotation approximation. Recent publications \cite{iloveq_aniso,ilovec_aniso,biswas} have extended this to anisotropic stars as well. We now examine if this universality can be extended to the magnetic, anisotropic NSs we have discussed so far in this paper.

\begin{figure}[!htpb]
	\centering
	\includegraphics[scale=0.3]{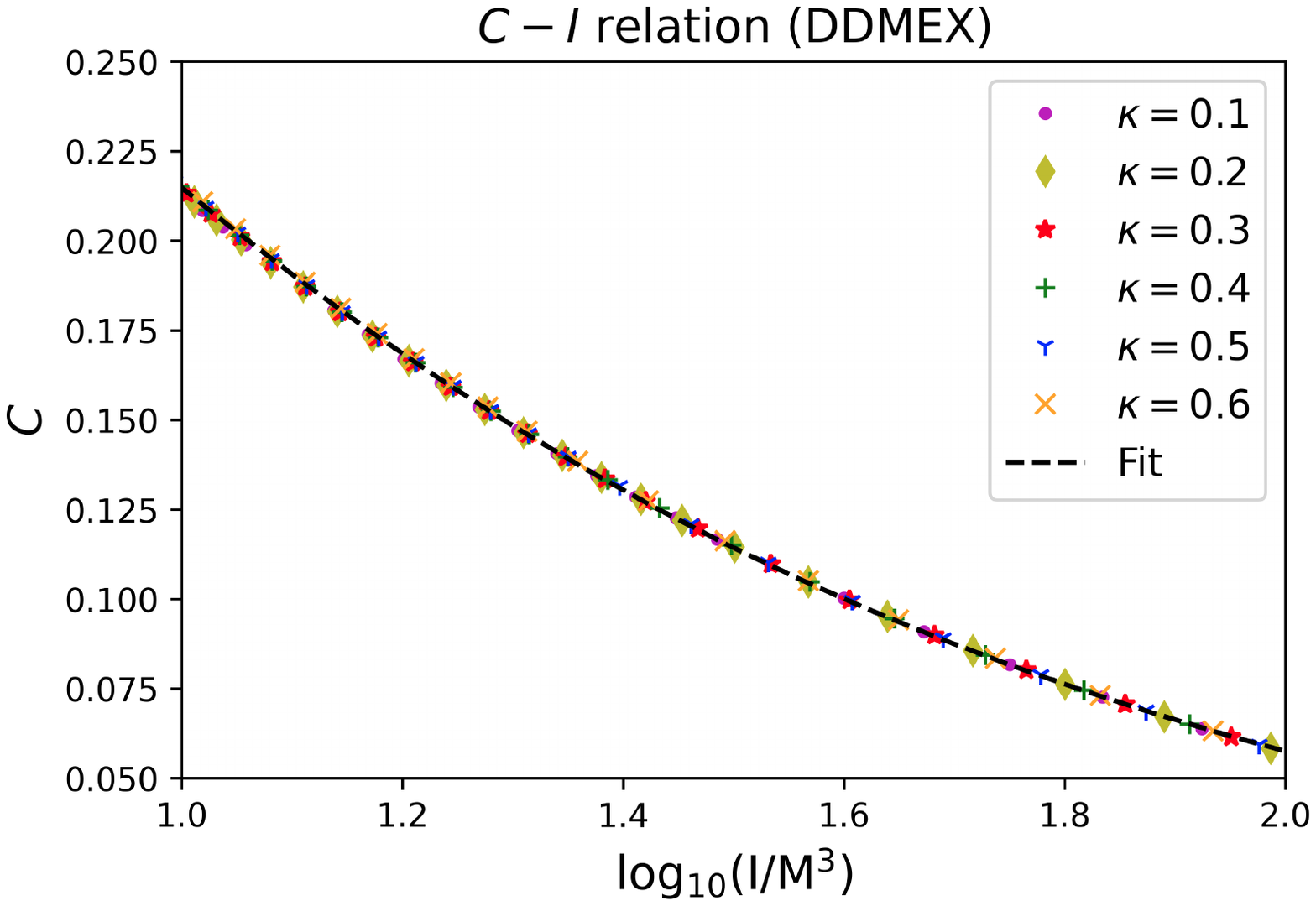}
	\caption{$C-I$ relation for nonmagnetic NSs constructed under DDMEX EOS with varying $\kappa$. $I$ is normalized in units of the cube of the NS's mass ($M^3$).}
	\label{CI_kappa}
\end{figure}

\begin{figure}[!htpb]
	\centering
	\includegraphics[scale=0.3]{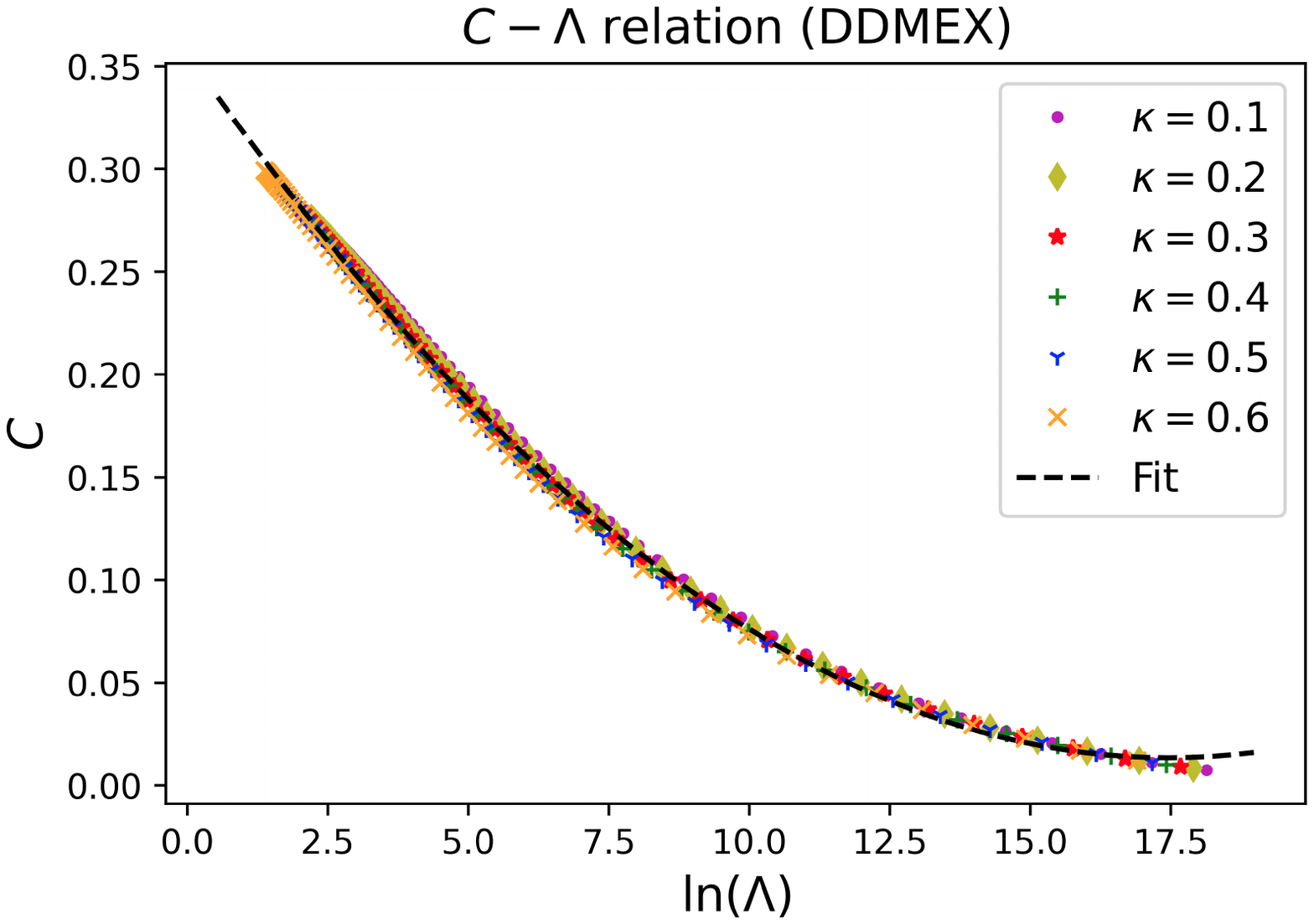}
	\caption{$C-\Lambda$ relation for nonmagnetic NSs constructed under DDMEX EOS with varying $\kappa$.}
	\label{CLambda_kappa}
\end{figure}

\begin{figure}[!htpb]
	\centering
	\includegraphics[scale=0.3]{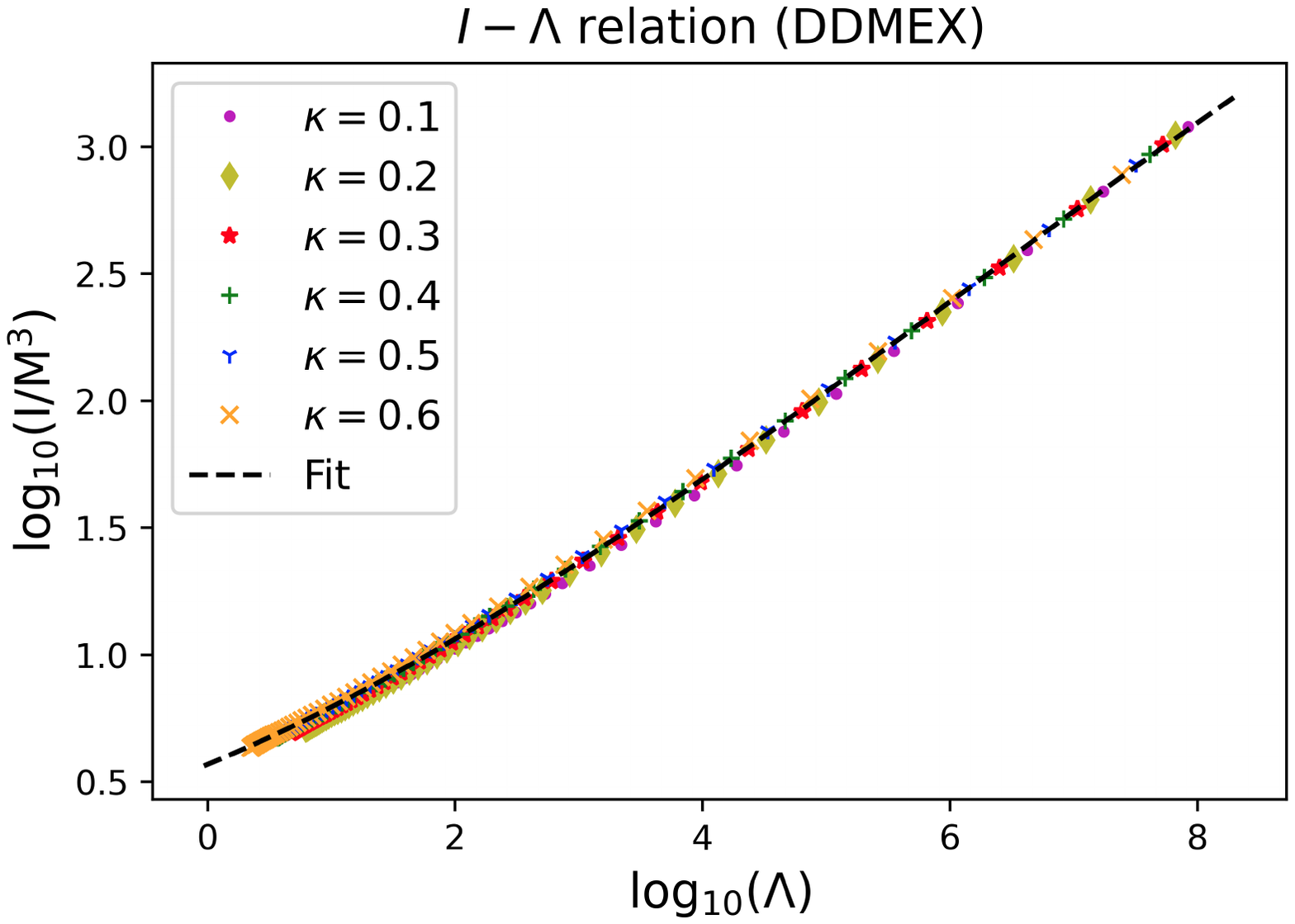}
	\caption{$I-\Lambda$ relation for nonmagnetic NSs constructed under DDMEX EOS with varying $\kappa$. $I$ is normalized in units of the cube of the NS's mass ($M^3$).}
	\label{Ilambda_kappa}
\end{figure}

In Figs. \ref{CI_kappa}, \ref{CLambda_kappa} and \ref{Ilambda_kappa}, the $C-I$, $C-\Lambda$ and $I-\Lambda$ relations are shown for the DDMEX EOS by varying $\kappa$ considering its different values as $0.1, \ 0.2, \ 0.3, \ 0.4, \ 0.5, \text{ and } 0.6$. As seen from the figures, the $C-I$ relations are unaffected by the presence of anisotropy in the star. However, the $C-\Lambda$ and $I-\Lambda$ relations show slight deviation once anisotropy is introduced, particularly at the low $\Lambda$ regime. This slight anisotropic deviation was also seen in previous work \cite{iloveq_aniso}. To find the extent of the deviation, we fit the relations using the least squares fitting.

We fit the $C-I$ relation using the following general polynomial \cite{I_uni,I_uni2},
\begin{equation}
    C = \sum_0^4 a_n(\text{log}_{10}I)^{-n},
    \label{CI_fitting}
\end{equation}
where $a_n$ represents the best-fit parameters.
For the $C-\Lambda$ relation fitting, we use the following general polynomial \cite{maselli},
\begin{equation}
    C = \sum_0^2 b_n(\text{ln}\Lambda)^{n},
    \label{CI_fitting}
\end{equation}
where $b_n$ represents the best-fit parameters.
Finally, for the $I-\Lambda$ relation we use \cite{I_uni,I_uni2},
\begin{equation}
    \text{log}_{10}I = \sum_0^3 c_n(\text{log}_{10}\Lambda)^{n},
    \label{CI_fitting}
\end{equation}
where $c_n$ represents the best-fit parameters. 

The calculated values of parameters $a_n, \ b_n, \text{ and } c_n$, along with the standard error (square root of variance) on each parameter are given in Table \ref{fit_DDMEX_K}. To quantify the goodness of fit, we calculate the root-mean-squared error (RMSE), given by the square root of the mean of the difference between the predicted and actual results squared. The corresponding RMSE for the varying $\kappa$ cases is given in Table \ref{RMSE_K}.

\begin{table}[!h]
\begin{tabular}{cccc}
\hline
                             & Fit parameter & Best-fit value           & Error                   \\ \hline
\multirow{5}{*}{$C-I$}       & $a_0$         & $-0.00891$                & $5.950 \times 10^{-3}$   \\  
                             & $a_1$         & $-0.2343$                 & $3.259 \times 10^{-2}$   \\ 
                             & $a_2$         & $1.1321$                & $6.236 \times 10^{-2}$   \\  
                             & $a_3$         & $-0.9169$                & $4.998 \times 10^{-2}$   \\  
                             & $a_4$         & $0.24284$                 & $1.432 \times 10^{-2}$   \\ \hline
\multirow{3}{*}{$C-\Lambda$} & $b_0$         & $0.35587$                & $5.983 \times 10^{-4}$  \\  
                             & $b_1$         & $-0.03921$               & $1.925 \times 10^{-4}$  \\  
                             & $b_2$         & $0.001123$                & $1.120 \times 10^{-5}$ \\ \hline
\multirow{4}{*}{$I-\Lambda$} & $c_0$         & $0.56707$                & $5.776 \times 10^{-3}$   \\  
                             & $c_1$         & $0.20232$               & $6.673 \times 10^{-3}$    \\  
                             & $c_2$         & $0.024963$               & $2.0170 \times 10^{-3}$     \\  
                             & $c_3$         & $-0.00134$              & $1.7265 \times 10^{-4}$             \\ \hline
\end{tabular}
\caption{The best-fit parameters and calculated errors for the universal relations for NSs constructed by DDMEX EOS with varying degrees of anisotropy.}
\label{fit_DDMEX_K}
\end{table}

\begin{table}[!htbp]
\begin{tabular}{cc}
\hline
Universal relation & RMSE ($\%$)       \\ \hline
$C-I$              & $0.1626$ \\ 
$C-\Lambda$        & $0.3721$ \\ 
$I-\Lambda$        & $1.9505$ \\ \hline
\end{tabular}
\caption{RMSE for the universal relations for NSs constructed by DDMEX EOS with varying degrees of anisotropy.}
\label{RMSE_K}
\end{table}

\begin{figure}[!htpb]
	\centering
	\includegraphics[scale=0.3]{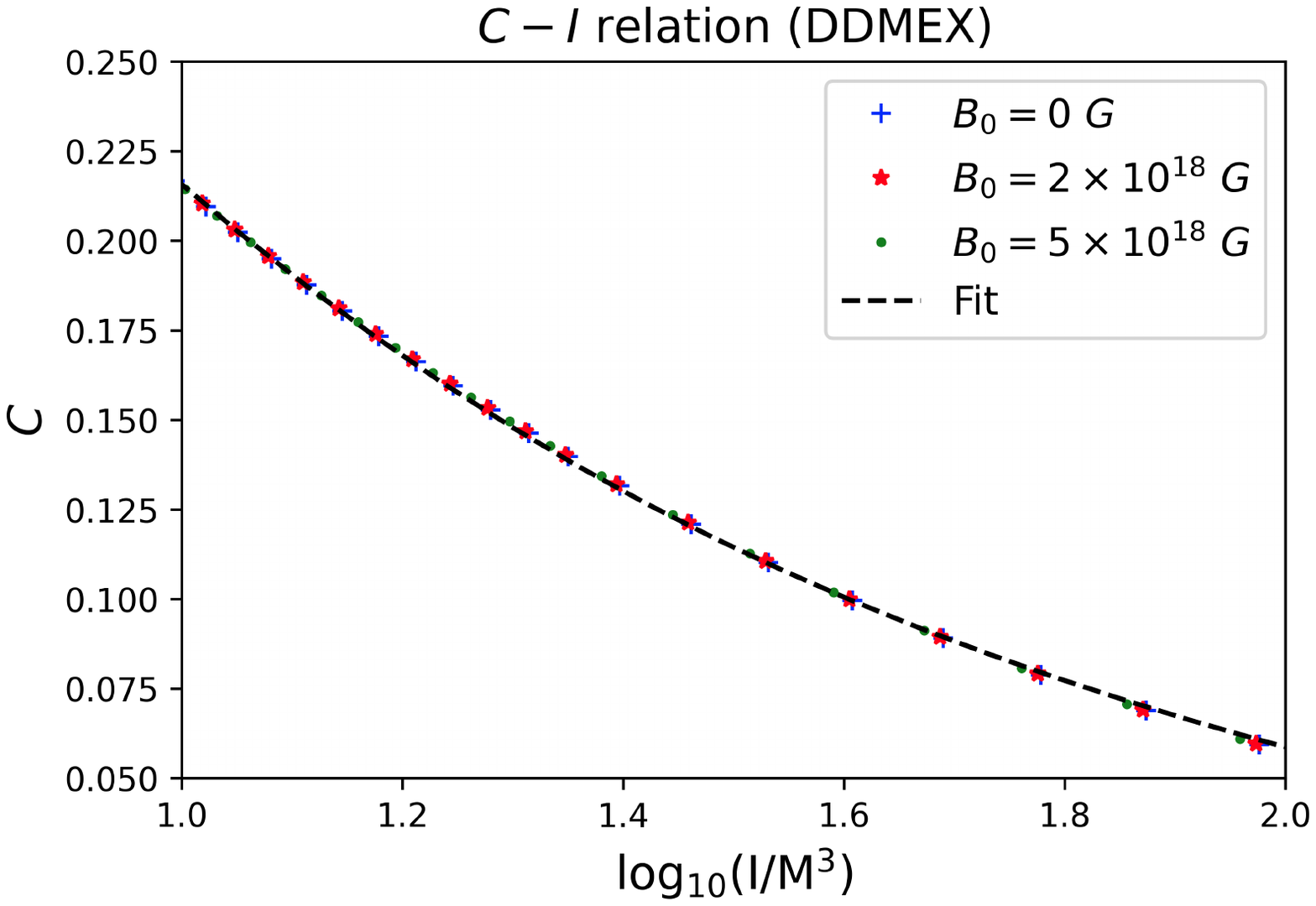}
	\caption{$C-I$ relation for DDMEX EOS with fixed $\kappa = 0.5$. Two values of $B_0$ are chosen along with the nonmagnetized case.}
	\label{CI_mag}
\end{figure}

\begin{figure}[!htpb]
	\centering
	\includegraphics[scale=0.3]{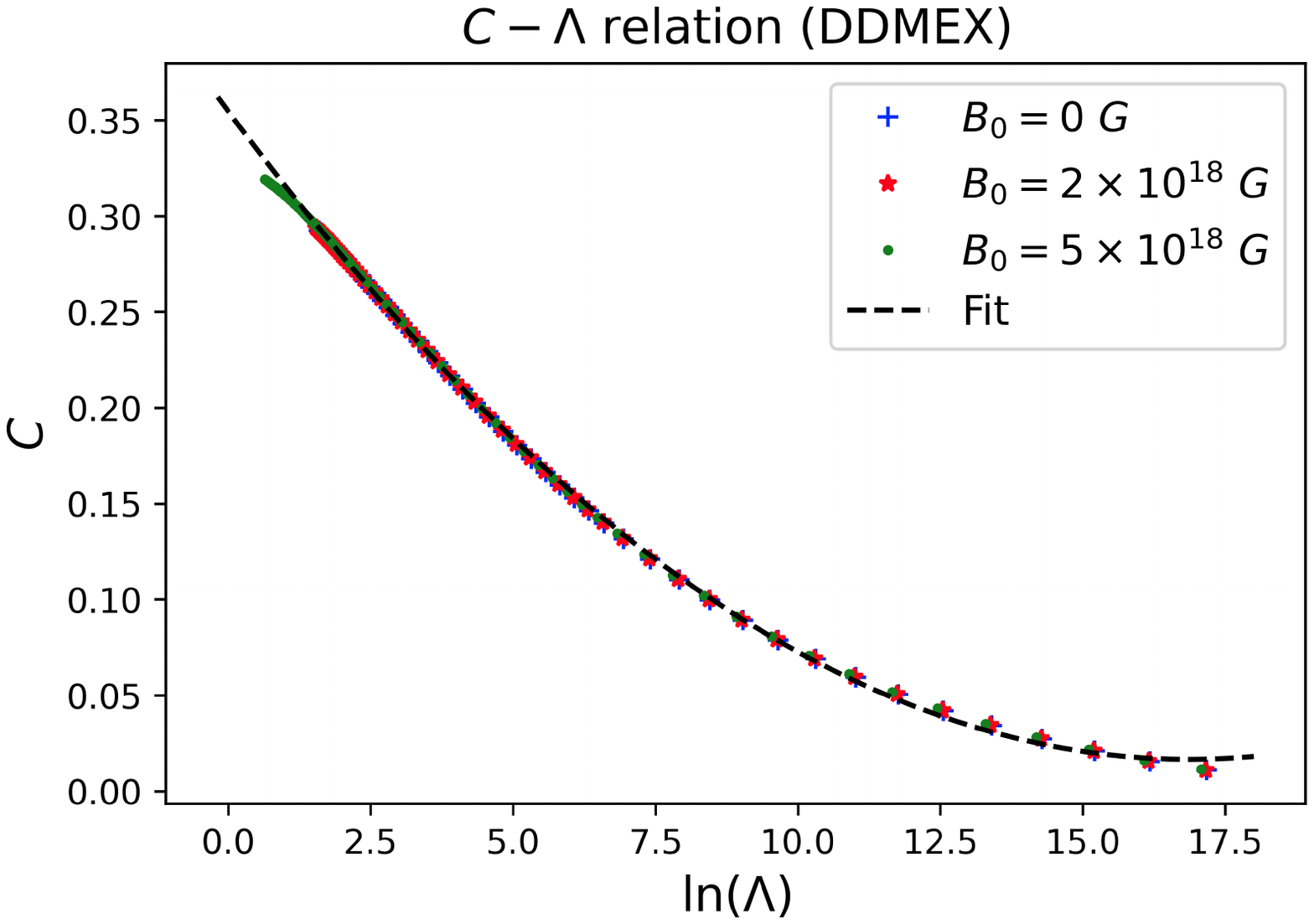}
	\caption{$C-\Lambda$ relation for DDMEX EOS with fixed $\kappa = 0.5$. Two values of $B_0$ are chosen along with the nonmagnetized case.}
	\label{CLambda_mag}
\end{figure}

\begin{figure}[!htpb]
	\centering
	\includegraphics[scale=0.3]{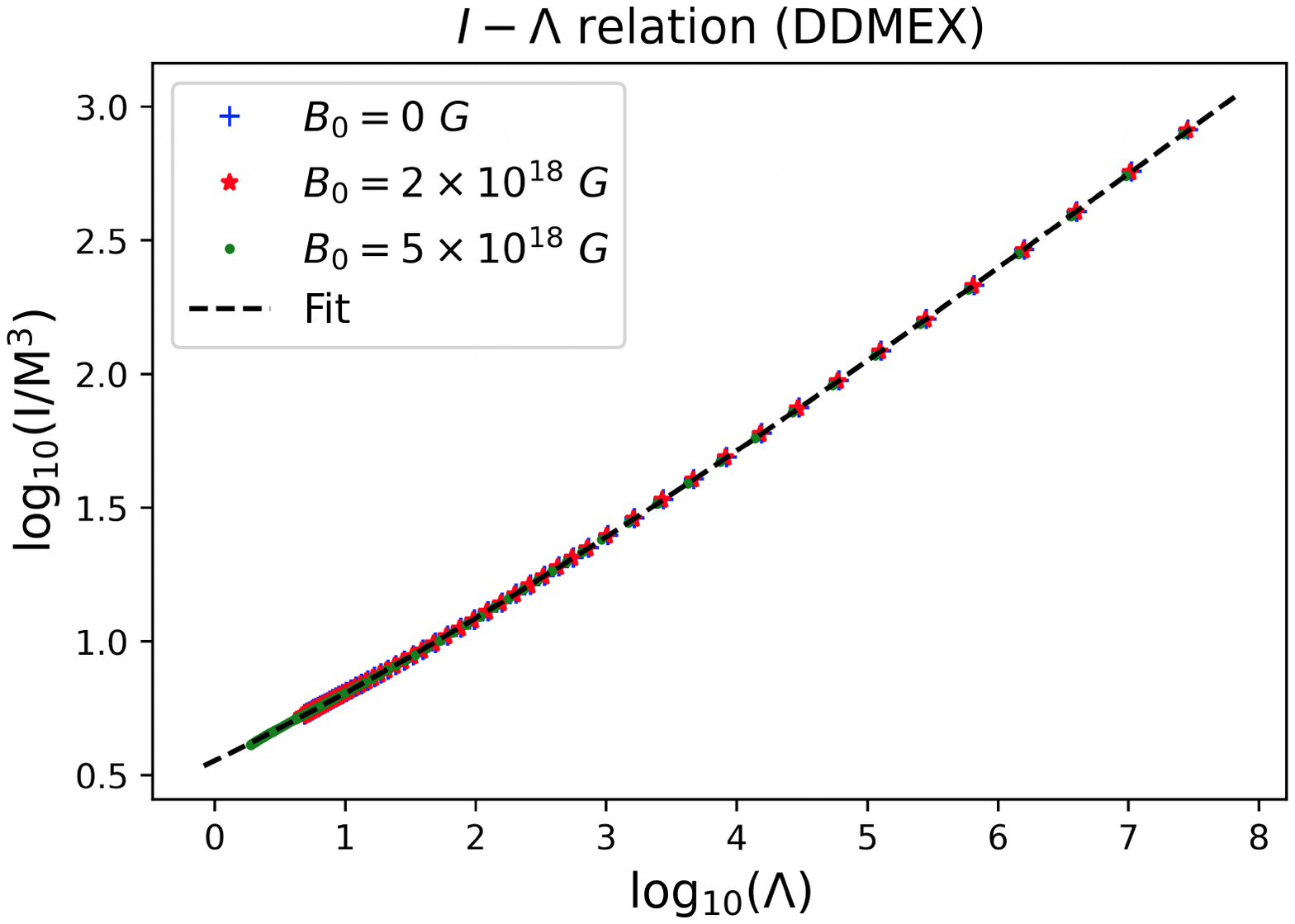}
	\caption{$I-\Lambda$ relation for DDMEX EOS with fixed $\kappa = 0.5$. Two values of $B_0$ are chosen along with the nonmagnetized case.}
	\label{Ilambda_mag}
\end{figure}

To examine the effect of magnetic field on the universal relations, we look at the same three relations - $C-I$, $C-\Lambda$, and $I-\Lambda$, but by fixing $\kappa$ to be $0.5$. Along with the nonmagnetized cases, we also include $B_0 = 2 \times 10^{18}$ and $B_0 = 5 \times 10^{18} \ {\rm G}$ cases. The results for the DDMEX EOS are shown in Figs. \ref{CI_mag}, \ref{CLambda_mag}, and \ref{Ilambda_mag}. We see that the universal relations are unaffected by the presence of the magnetic field.

The fitting is done as explained previously. The best-fit parameters, along with the errors on the same, are given in Table \ref{fit_mag_DDMEX}. The RMSE is also listed in Table \ref{RMSE_Mag}.

\begin{table}[!htbp]
\begin{tabular}{cccc}
\hline
                             & Fit parameter & Best-fit value           & Error                    \\ \hline
\multirow{5}{*}{$C-I$}       & $a_0$         & $-0.06922$               & $2.751 \times 10^{-3}$   \\ 
                             & $a_1$         & $0.11608$                & $1.362 \times 10^{-2}$  \\  
                             & $a_2$         & $0.42245$                & $2.322 \times 10^{-2}$   \\ 
                             & $a_3$         & $-0.31825$               & $1.639 \times 10^{-2}$   \\  
                             & $a_4$         & $0.064622$                & $4.100 \times 10^{-3}$   \\ \hline
\multirow{3}{*}{$C-\Lambda$} & $b_0$         & $0.35466$                & $3.4206 \times 10^{-4}$  \\
                             & $b_1$         & $-0.04009$              & $1.1658 \times 10^{-4}$  \\  
                             & $b_2$         & $0.00118$                & $7.0507 \times 10^{-6}$ \\ \hline
\multirow{4}{*}{$I-\Lambda$} & $c_0$         & $0.5531$                & $6.0172 \times 10^{-4}$  \\ 
                             & $c_1$         & $0.2381$                & $8.5816 \times 10^{-4}$  \\ 
                             & $c_2$         & $0.015847$               & $2.94299 \times 10^{-4}$ \\ 
                             & $c_3$         & $-0.000723$              & $2.7440 \times 10^{-5}$ \\ \hline
\end{tabular}
\caption{The best-fit parameters and calculated errors for the universal relations for NSs constructed by DDMEX EOS with fixed $\kappa = 0.5$ and varying $B_0$.}
\label{fit_mag_DDMEX}
\end{table}

\begin{table}[!htbp]
\begin{tabular}{cc}
\hline
Universal relation & RMSE ($\%$)      \\ \hline
$C-I$              & $0.0745$ \\ 
$C-\Lambda$        & $0.1574$ \\ 
$I-\Lambda$        & $0.2402$ \\ \hline
\end{tabular}
\caption{RMSE for the universal relations of NSs constructed by DDMEX EOS with fixed $\kappa = 0.5$ and varying $B_0$.}
\label{RMSE_Mag}
\end{table}

\begin{figure}[!htbp]
	\centering
	\includegraphics[scale=0.3]{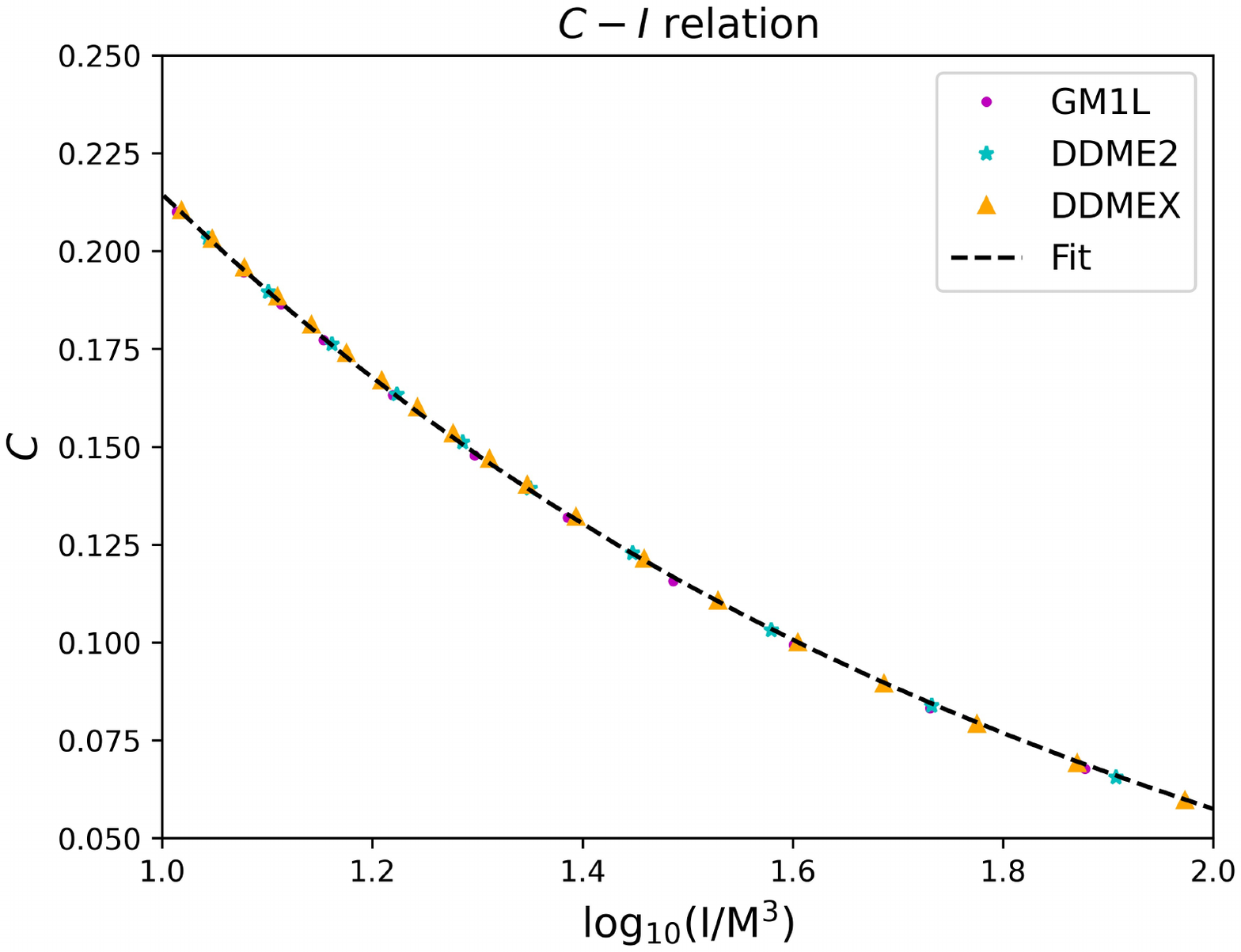}
	\caption{$C-I$ relation for NSs constructed by DDMEX, GM1L, and DD-ME2 EOS with fixed $\kappa = 0.5$ and fixed $B_0 = 2 \times 10^{18} \ {\rm G}$.}
	\label{CI_EOS}
\end{figure}

\begin{figure}[!htpb]
	\centering
	\includegraphics[scale=0.3]{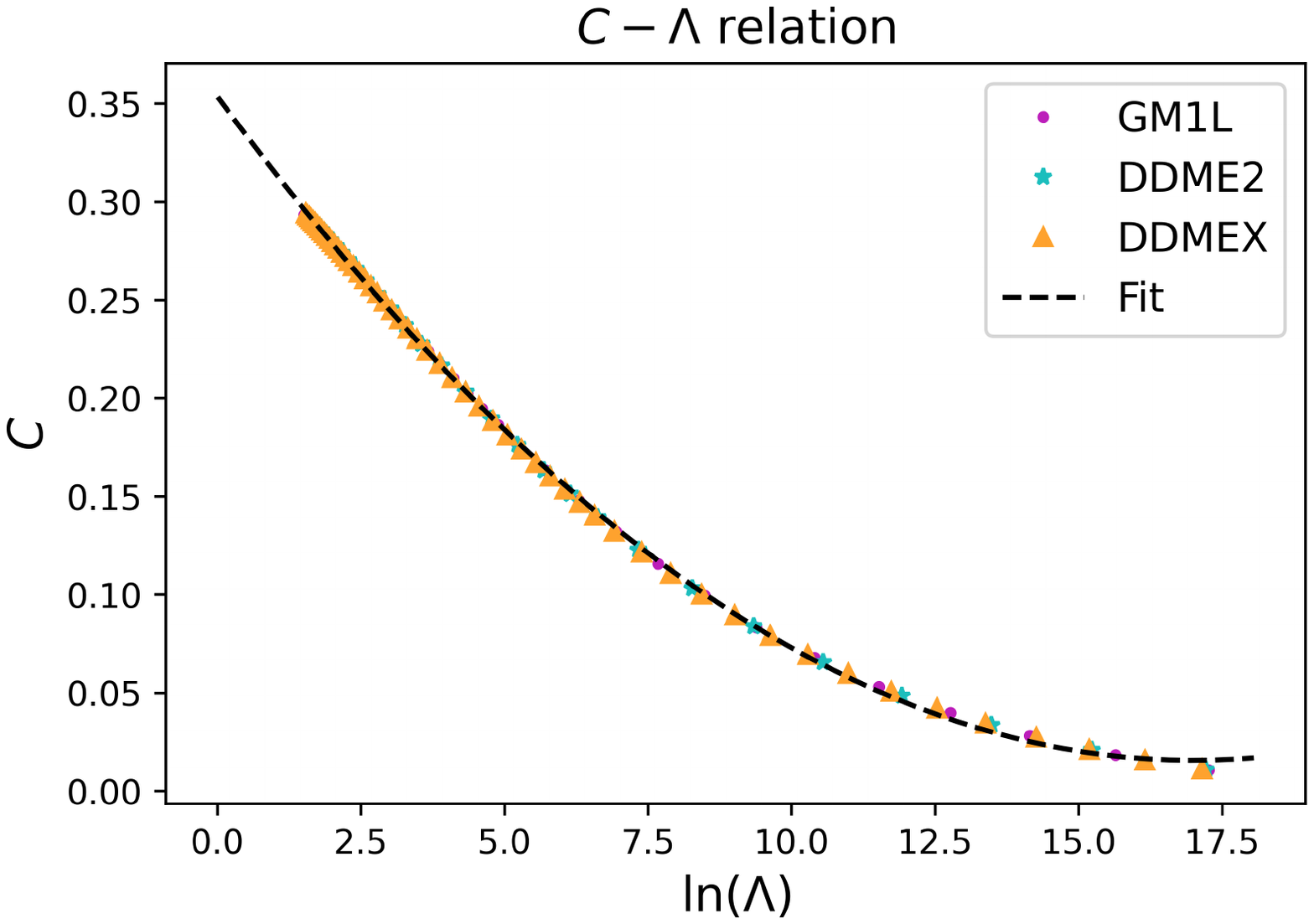}
	\caption{$C-\Lambda$ relation for NSs constructed by DDMEX, GM1L, and DD-ME2 EOS with fixed $\kappa = 0.5$ and fixed $B_0 = 2 \times 10^{18} \ {\rm G}$.}
	\label{CLambda_EOS}
\end{figure}

\begin{figure}[!htpb]
	\centering
	\includegraphics[scale=0.3]{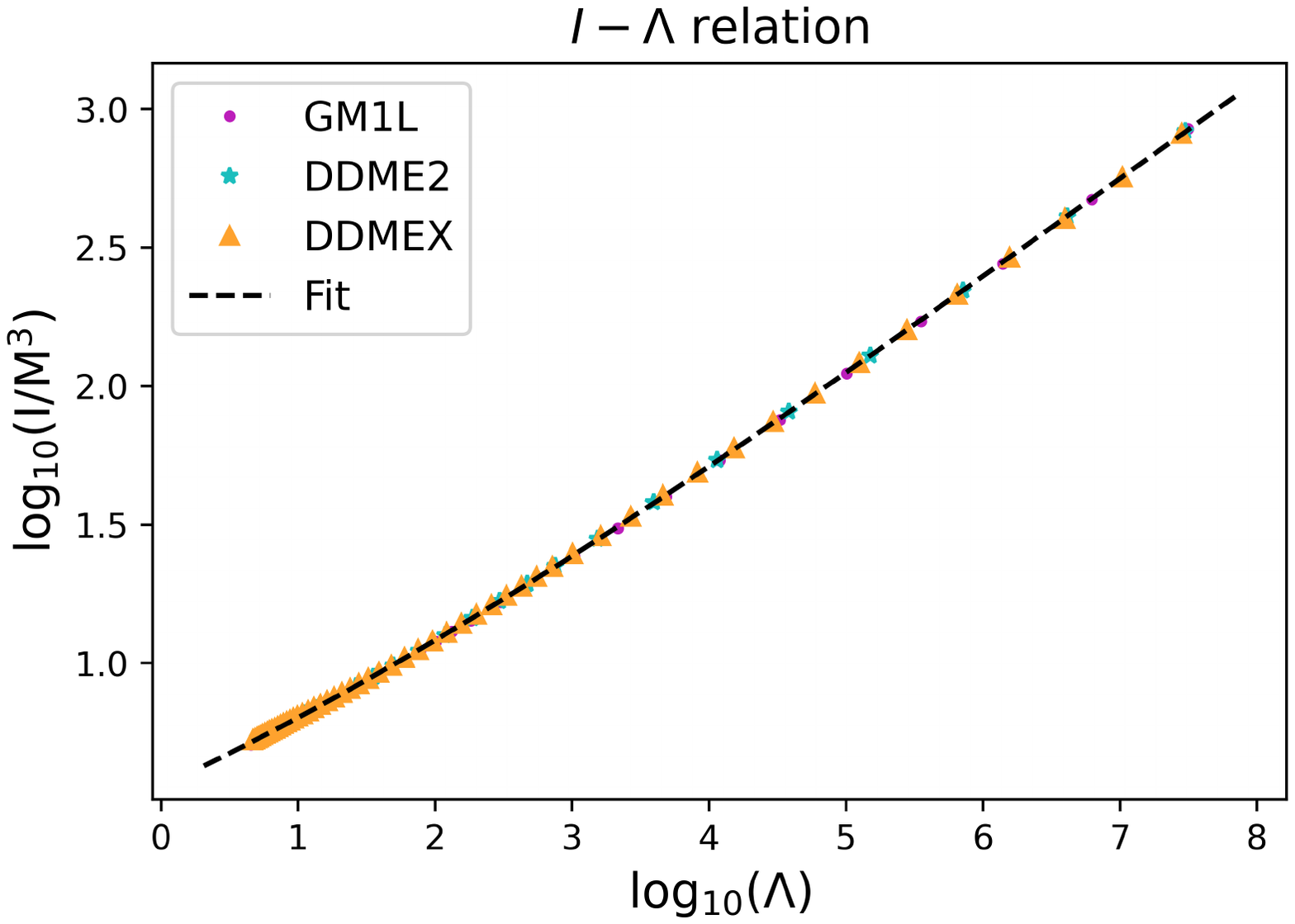}
	\caption{$I-\Lambda$ relation for NSs constructed by DDMEX, GM1L, and DD-ME2 EOS with fixed $\kappa = 0.5$ and fixed $B_0 = 2 \times 10^{18} \ {\rm G}$.}
	\label{Ilambda_EOS}
\end{figure}

Finally, we check the universality of the $I$-Love-$C$ relations across EOS. In Figs. \ref{CI_EOS}, \ref{CLambda_EOS}, and \ref{Ilambda_EOS}, each of the universal relations are shown for NSs for three different EOS: GM1L, DD-ME2, and DDMEX. We fix the anisotropy parameter $\kappa$ = 0.5 and magnetic field with $B_0 = 2 \times 10^{18} \ {\rm G}$. We see that the universal relations remain preserved across EOS for magnetized stars with fixed anisotropy. In each figure, the fitting is done as explained previously. The best-fit parameter values and the errors on them are listed in Table \ref{eos_fit}. The RMSE for the fits are shown in Table \ref{RMSE_EOS}.

\begin{table}[!htbp]
\begin{tabular}{cccc}
\hline
                             & Fit parameter & Best-fit value           & Error                    \\ \hline
\multirow{5}{*}{$C-I$}       & $a_0$         & $-0.1127$               & $2.309 \times 10^{-2}$ \\  
                             & $a_1$         & $0.29967$                 & $1.106 \times 10^{-1}$  \\ 
                             & $a_2$         & $0.14764$               & $1.900 \times 10^{-1}$  \\ 
                             & $a_3$         & $-0.1451$                & $1.395 \times 10^{-1}$  \\ 
                             & $a_4$         & $0.02503$               & $3.371 \times 10^{-2}$  \\ \hline
\multirow{3}{*}{$C-\Lambda$} & $b_0$         & $0.35364$                & $4.184 \times 10^{-4}$  \\
                             & $b_1$         & $-0.03977$              & $1.412 \times 10^{-4}$  \\ 
                             & $b_2$         & $0.00117$               & $8.420 \times 10^{-6}$ \\ \hline
\multirow{4}{*}{$I-\Lambda$} & $c_0$         & $0.55371$                & $2.248 \times 10^{-3}$ \\  
                             & $c_1$         & $0.2322$               & $2.768 \times 10^{-3}$ \\ 
                             & $c_2$         & $0.017612$              & $8.679 \times 10^{-4}$ \\ 
                             & $c_3$         & $-0.0008565$               & $7.622 \times 10^{-5}$ \\ \hline
\end{tabular}
\caption{The best-fit parameters and calculated errors for the universal relations of NSs constructed by DDMEX, GM1L, and DD-ME2 EOS with fixed $\kappa = 0.5$ and fixed $B_0 = 2 \times 10^{18} \ {\rm G}$.}
\label{eos_fit}
\end{table}

\begin{table}[!htbp]
\begin{tabular}{cc}
\hline
Universal relation & RMSE ($\%$)       \\ \hline
$C-I$              & $0.1629$ \\
$C-\Lambda$        & $0.1636$ \\ 
$I-\Lambda$        & $0.4941$ \\ \hline
\end{tabular}
\caption{RMSE for the universal relations of NSs constructed by DDMEX, GM1L, and DD-ME2 EOS with fixed $\kappa = 0.5$ and fixed $B_0 = 2 \times 10^{18} \ {\rm G}$.}
\label{RMSE_EOS}
\end{table}

Thus, we find that under the model anisotropy and magnetic field considered, the universal relations $I$-Love-$C$ are mostly preserved. It appears that of all effects considered, it is the anisotropy that has a maximal effect on the universality of these relations, more precisely on $C-\Lambda$ and $I-\Lambda$ relations. Varying $\kappa$ makes the RMSE of the fit roughly 2-4 times that of the case with fixed $\kappa$. However, varying magnetic field while keeping $\kappa$ fixed gives, indeed, a better universal fit. As explained in previous work \cite{biswas}, the rigorous exploration of the effect of anisotropy on the $I$-Love-$C$ relations could lead to possible observational constraints on $\kappa$ and the degree of anisotropy in the star in general.


\section{Comparison between polynomial and exponential profiles}
\label{sec_dex}

All the above explorations are based on a particular law of the density-dependent magnetic field profile, which offers an exponential variation. 
One may question if the density-dependent magnetic field profile used in this work (Eq. \ref{mag_bando}) is inconsistent with the Maxwell equations \cite{mag_crit}. Actually, this should not be an issue. Under the formalism of approximate spherical symmetry (which is quite a good approximation, particularly, for toroidal fields) described in this paper, the density-dependent profile only gives us the magnitude of the magnetic field at a particular density (and hence, radius) within the star. The orientation is determined by whether we consider ``RO" or ``TO" fields. As we have shown in the previous work in the same line, we can always make the magnitude obtained from the density-dependent profile consistent with the Maxwell equations, approximately in the regime of spherical symmetry, for both of the orientations considered in this work. A detailed proof of this is shown in Sec. 2.4 of \cite{deb}, under similar formalism and assumptions.

One of the alternative profiles is a polynomial profile, with magnetic field expressed as a function of the baryon chemical potential ($\mu_B$). The profile is obtained by fitting the two-dimensional results from the numerical code LORENE using quadratic polynomials \cite{dex1,dex2}, given by
\begin{equation}
    \label{dex}
    B(\mu_B) = \frac{a + b\mu_B + c\mu_B^2}{B_c^2}\mu,
\end{equation}
where $\mu_B$ is the baryon chemical potential and $\mu$ is the dipole magnetic moment, respectively, expressed in MeV and $A/m^2$, to obtain $B$ in units of the electron's critical field $B_c = 4.414 \times 10^{13} \ {\rm G}$. We use coefficients, $a,~b,~c$, obtained from the numerical simulation of a $2.2 M_\odot$ star in previous work \cite{dex1,dex2}, which are $a = -0.769 \ {\rm G}^2/{\rm Am}^2, \ b = 1.20 \times 10^3 \ {\rm G}^2/{\rm Am}^2\ {\rm MeV}, \ c = - 3.46 \times 10^{-7} \ {\rm G}^2/{\rm Am}^2\ {\rm MeV}^2$. 
As implied, this profile, in principle, is valid only for $2.2 M_\odot$ NS.
Other NSs should have different $a,~b,~c$. 

We introduce magnetic fields to our star using the above profile by varying $\mu$ with values $2 \times 10^{31}, 5 \times 10^{31}, 10^{32}, {\rm and}~\ 2 \times 10^{32} \ {\rm Am}^2$. An important caveat, however, here is that LORENE can only generate poloidal magnetic fields. In the current formalism, this translates to an ``RO" field. We set the anisotropy $\kappa = 0.1$ throughout.

\begin{figure}[!htpb]
	\centering
	\includegraphics[scale=0.3]{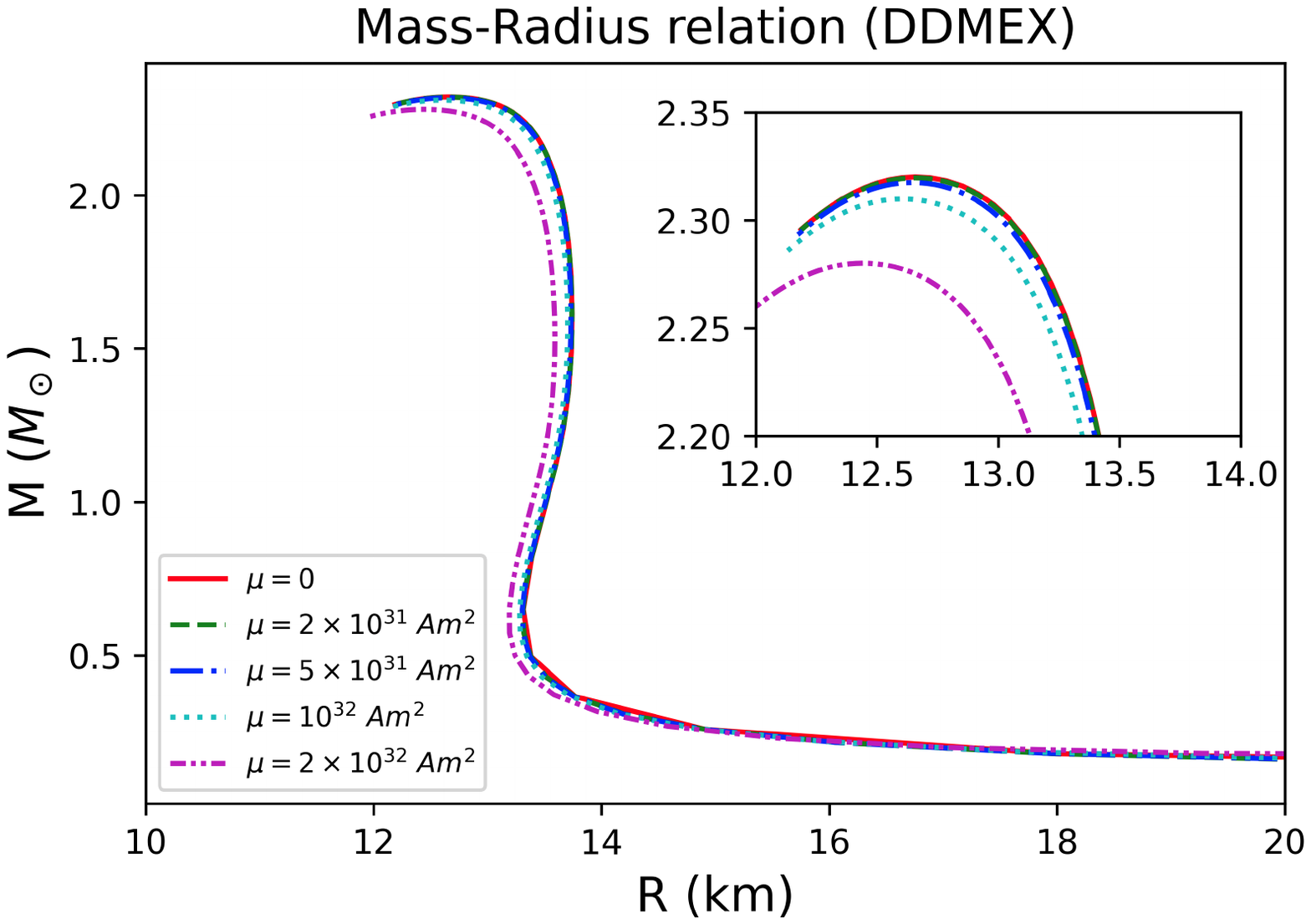}
	\caption{M-R curves for varying dipole magnetic moment $\mu$ for the DDMEX EOS, where $\kappa=0.1$ throughout.}
	\label{dex_mr_curves}
\end{figure}

\begin{figure}[!htpb]
	\centering
	\includegraphics[scale=0.3]{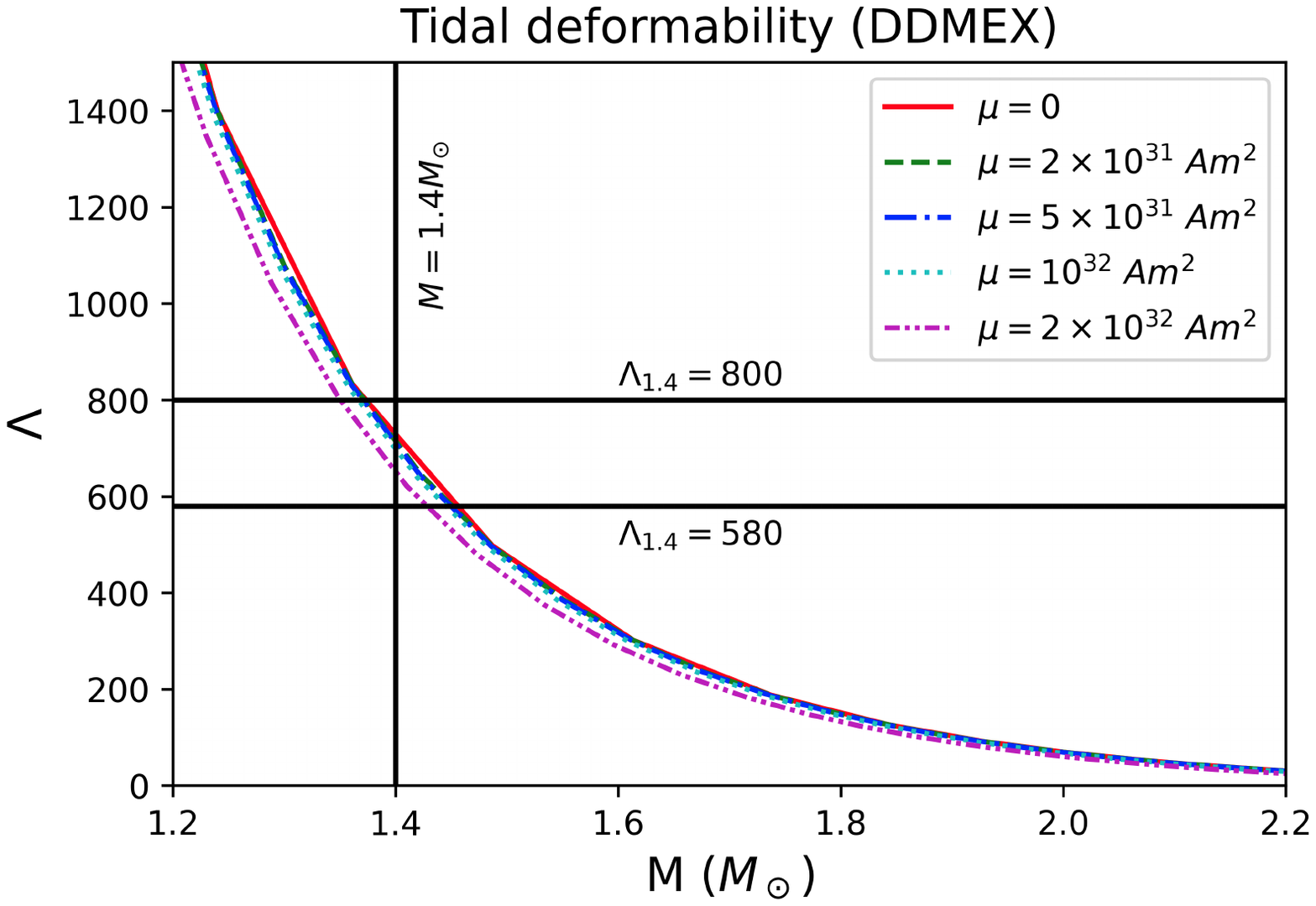}
	\caption{Tidal deformability $\Lambda$ as a function of M for varying dipole magnetic moment $\mu$ for the DDMEX EOS, where $\kappa=0.1$ throughout.}
	\label{dex_tidal}
\end{figure}

The M-R relations for the four magnetized cases mentioned above, along with a nonmagnetized (but still anisotropic) case for the DDMEX EOS are shown in Fig. \ref{dex_mr_curves}. The corresponding values of the tidal deformability are shown in Fig. \ref{dex_tidal}. As this profile leads to pure poloidal/RO fields, we see that under the current formalism, it leads to an overall decrease in the mass of the NS. 

\begin{figure}[!htpb]
	\centering
	\includegraphics[scale=0.3]{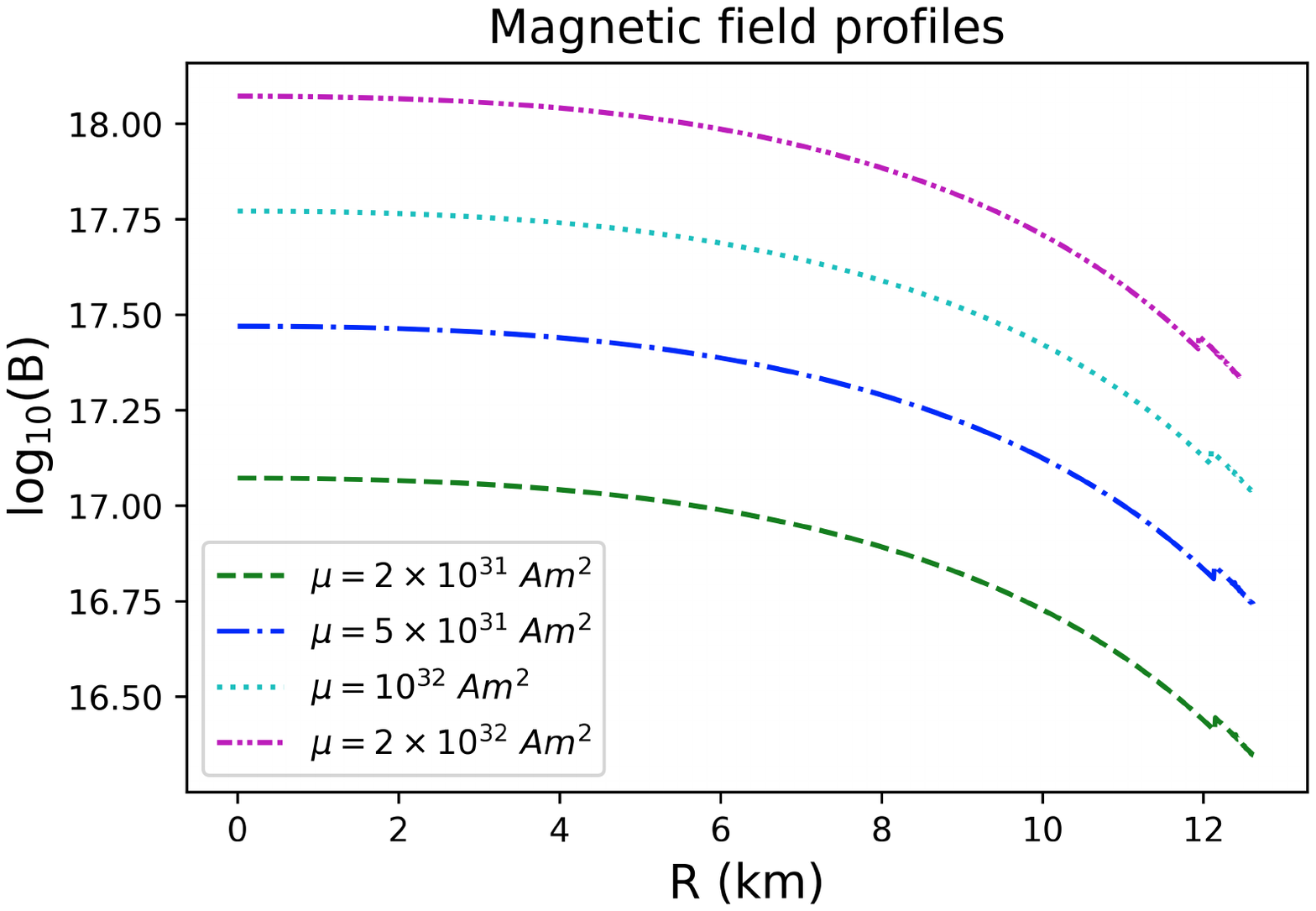}
	\caption{Variation of the magnetic field within the star for varying $\mu$ based on the polynomial profile with the DDMEX EOS.}
	\label{dex_profiles}
\end{figure}

It appears that the magnetic field effect is extremely limited in the case of this profile as opposed to the previous profile. From the results listed in Table \ref{results_dex}, we see that the NS properties are not all that different from the nonmagnetized cases even for high magnitudes of the central (maximum) magnetic field and $E_{mag}/E_{grav}$. The trend of variation of the magnetic field within the star for each of the $\mu$ cases is shown in Fig. \ref{dex_profiles}. We see that this profile always gives, at most, an order-of-magnitude difference between the center and the surface fields. This means that a maximum magnetic field of order $10^{18} \ {\rm G}$ within the star corresponds to a surface field of similar order or, at the maximum, reduced up to $10^{17} \ {\rm G}$. This is much higher than the present measurements for the surface magnetic field of NSs. Additionally, such a high field throughout the star leads to high $E_{mag}/E_{grav}$ (Table \ref{results_dex}), which might destabilize the star. On the other hand, if we aim to achieve the surface field $\sim 10^{15-16}$ G, then the corresponding central field hardly affects the stellar structure. Therefore, this profile does not seem to be a viable profile, from either stability or observability.

\begin{table}[!htbp]
\begin{tabular}{ccccc}
\hline
$\mu \ ({\rm Am}^2)$     & $B_{c} \ ({\rm G})$         & $\rm{M}_{max}$ $(M_\odot)$ & R (km) & $E_{mag}/E_{grav}$ \\ \hline
0                  & -                     & 2.32                & 12.66     & -                  \\ 
$2 \times 10^{31}$ & $1.18 \times 10^{17}$ & 2.32                & 12.65     & 0.002             \\ 
$5 \times 10^{31}$ & $2.95 \times 10^{17}$ & 2.32                & 12.64     & 0.01              \\ 
$10^{32}$          & $5.90 \times 10^{17}$ & 2.31               & 12.60     & 0.04             \\ 
$2 \times 10^{32}$ & $1.18 \times 10^{18}$ & 2.28                & 12.44     & 0.17             \\ \hline
\end{tabular}
\caption{Results for the DDMEX EOS with varying $\mu$ in the polynomial magnetic field profile, where $\kappa=0.1$ throughout.}
\label{results_dex}
\end{table}

Indeed, this 1 order-of-magnitude variation of the magnetic field from the center to surface is consistent with previous work, done using this profile \cite{kundu_sinha,dex1,dex2}. It is also consistent with the numerical simulation results for the poloidal fields from our group using the XNS code \cite{mayu, surajit}. However, it has been long known \cite{tayler1,tayler2} that purely poloidal or purely toroidal field configurations lead to unstable stars. The most stable field configuration in NSs is a mixed field configuration consisting of both poloidal and toroidal fields; perhaps toroidally dominated \cite{wickramasinghe}, also see \cite{landerjones, pili}. Hence, it would be incorrect to rule out the magnetic field's effect on the star based on the results from the purely poloidal polynomial profile. In fact, XNS code shows that toroidal fields can have 2 orders of magnitude variation between the maximum field and the outer crust field (exactly at the surface it exhibits zero or very low field by the very geometry). Hence, the chosen field profile with the exponential variation, as used in all the previous sections, is likely to be a natural situation with the right combination of toroidal and poloidal fields within the star. The toroidal field varies from its maximum to minimum (close to the surface) by 2 orders of magnitude and the corresponding poloidal field varies by one order. If the core field is toroidally dominated with 2 orders of magnitude higher than its poloidal counterpart, then for a maximum toroidal field of $10^{18}$ G, the surface poloidal field could be $10^{15}$ G (exactly at the surface, the toroidal field vanishes). In such a star, the poloidal field hardly affects the stellar structure. This further justifies the approximate spherical choice of our model, when a significant deviation from the spherical symmetry is only by the poloidal field. 

Thus, although the polynomial profile is constructed in order to be consistent with the Einstein-Maxwell equations, it does not account for the instabilities that arise from purely poloidal configurations as well as possible instability due to high $E_{mag}/E_{gav}$. Indeed, all the profile-based approaches, whether they be the exponential profile or the polynomial profile, are ultimately approximations that have limitations in predicting the NS physics. For a first-order study into the NS's mass limits, such as the one done in this work, it appears that the exponential profile is more than adequate. For more sophisticated studies of the magnetic field profiles and the precise effect of its geometry, one must turn to two-dimensional Einstein-Maxwell solvers such as XNS, as done already by some of us \cite{surajit, mayu}. However, at present, the XNS code cannot deal with realistic EOS such as the ones studied in this paper, it can deal only with the polytropic form. Thus the approximate studies, as reported in the present paper, become important to gain insights into the EOS dependence of NS properties.

\section{conclusions}\label{sec_con}

In this work, we investigate the possibility of massive NSs being mass gap candidates to explain observations such as GW190814. We first examine the pure EOS effect. Because of the appearance of exotic particles such as hyperons, the EOS is softened considerably as compared to the pure nucleonic case. The maximum mass obtained for isotropic stars constructed from our chosen set of $npe\mu-Y\Delta$ EOS was only $\simeq 2.2 M_\odot$. To truly bring our theoretical models to mass gap ranges, we next investigate the effects of magnetic field and a model anisotropy in the NS. Crucially, both magnetic field and anisotropy are well-established physical effects that show up in NSs. The observations of magnetars, for example, support NSs having surface fields as high as $10^{15} \ {\rm G}$. Anisotropy is also expected to be present, both as a consequence of the magnetic field introduced, and as an independent effect arising from high-density effects such as superfluidity. Introducing anisotropy enhances the mass of the star, even in the absence of the magnetic field. Depending on the orientation of the magnetic field introduced (RO or TO), the maximum mass can be either enhanced or decreased from the nonmagnetized value. Only the toroidal field leads to a reasonable increment in the mass of a NS. Although we introduce these two nonspherical effects to the system, we construct our NS models in approximate spherical symmetry. This is a reasonable approximation, particularly for toroidally dominated stars, as shown previously by our group using the two-dimensional numerical simulation code XNS \cite{surajit}. 

Nevertheless, many authors (e.g. \cite{dex_prc, li, tu}) tackle the EOS based restriction to the NS mass through the introduction of additional effects such as rotation in the star. There have also been arguments based on GW170817 observations combined with universal relations that the maximum mass of non-rotating, nonmagnetized NSs is within $2.3M_\odot$ \cite{rezzolla}; see also \cite{thapa_particles}. However, the hyperonization/softening issue may be confronted through the adjustment of the coupling constant as well \cite{lopes_GW190814}, which though appears to be an extreme assumption \cite{Sedrakian:2020PRD}. Moreover, the possibility of the ligher component of GW190814 to be a quark star is not ruled out \cite{han_steiner}. Therefore, while the alternative choices may lead to slightly varied outcomes, the fundamental findings of the paper persist in their robustness.

We further use $\Lambda$ and $E_{mag}/E_{grav}$ to ensure the physicality of our results. We find that high magnetic fields from certain profiles are ruled out under these constraints. However, with the right choice of profile, NSs of masses in the range $2.5 - 2.67 M_\odot$ are demonstrated to be possible, while satisfying all stability criteria. On examining the role of anisotropy in the star, we find that increasing the anisotropy parameter $\kappa$ leads to a decrease in the tidal deformability. This is due to the enhanced mass, as the tidal deformability is inversely proportional to the compactness. This means that EOS that were previously ruled out based on isotropic studies of the $\Lambda_{1.4}$ constraint are still viable candidates if we consider their anisotropic pressure modification. The sensitivity of $\Lambda$ on $\kappa$ is extremely important as it may help to ultimately establish observational bounds on this parameter. 

We finally examine the EOS independence of the $I$-Love-$C$ universal relations. We see that of the three effects - EOS, magnetic field, and anisotropy - it is the anisotropy that has a maximal effect of varying the universal fit between the parameters. Nevertheless, the deviation from universality is always with RMSE of the order of $1 \%$, which is still much smaller than experimental errors on, say, the moment of inertia $I$.

In summary, NSs could possibly be mass-gap candidates. However, the EOS alone is unlikely to yield a massive NS with a mass exceeding 
$2.5M_\odot$.
The introduction of anisotropy, arising from either matter, magnetic fields, or a combination of both, appears to be a crucial factor in achieving this. However, 
the magnetic field geometry must be carefully controlled to prevent instability and comply with the constraint imposed by tidal deformability.

With the appropriate magnetic field geometry and strength, it becomes
feasible to have a massive, stable NS that fits within the
mass-gap while satisfying the tidal deformability constraint. It is
important to note that we have not considered the effects of rotation.
Building on the findings of previous work \cite{Sedrakian:2020PRD}, an additional increase in mass will occur, which
should facilitate the interpretation of compact stellar objects within
the mass gap as NSs rather than low-mass black holes. This
aspect will be a subject of investigation in future research.

\section*{ACKNOWLEDGEMENTS}
Z. Z. and B. M. acknowledge Debabrata Deb of IMSc for the discussion about tidal deformability, Surajit Kalita of the University of Cape Town for the discussion about the variation of magnetic field profile, and Somnath Mukhopadhyay of NIT-Trichy for comments. Grateful thanks are due to the referee as well, who reported very promptly with constructive suggestions, which helped to improve the paper. Z. Z. acknowledges the Prime Minister’s Research Fellows (PMRF) scheme, with Ref. No. TF/PMRF-22-7307, for providing fellowship. B. M. acknowledges a project funded by SERB, India, with Ref. No. CRG/2022/003460, for partial support toward this research. F. W. is supported by the U.S. National Science Foundation under Grant No. PHY-2012152.

\bibliographystyle{apsrev4-1}
\bibliography{biblio}

\end{document}